\documentclass[12pt]{article}
\usepackage{epsfig, epsf, graphicx, float, color}
\usepackage{pstricks, psfrag}
\usepackage{amssymb,amsmath,mathrsfs}
\usepackage{verbatim,enumerate}
\usepackage{setspace}
\usepackage{natbib}

\usepackage{tikz-qtree,tikz-qtree-compat}

\usepackage{tikz}
\usepackage{mathpazo}
\usetikzlibrary{positioning}

\def\diag{\hbox{diag}}

\def\diag{\hbox{diag}}

\def\boxit#1{\vbox{\hrule\hbox{\vrule\kern6pt
          \vbox{\kern6pt#1\kern6pt}\kern6pt\vrule}\hrule}}

\def\cov{\hbox{cov}}

\def\bse{\begin{eqnarray*}}
\def\ese{\end{eqnarray*}}
\def\be{\begin{eqnarray}}
\def\ee{\end{eqnarray}}
\def\bq{\begin{equation}}
\def\eq{\end{equation}}
\def\bse{\begin{eqnarray*}}
\def\ese{\end{eqnarray*}}

\def\T{^{\rm T}}

\newcommand{\corb}[1]{\textcolor{black}{#1}}
\newcommand{\corred}[1]{\textcolor{black}{#1}}

\newcommand{\bbR}{\mathbb{R}}

\newcommand{\bbN}{\mathbb{N}}
\newcommand{\bbE}{\mathbb{E}}
\newcommand{\bbS}{\mathbb{S}}
\newcommand{\bD}{\mathbf{D}}
\newcommand{\bI}{\mathbf{I}}
\newcommand{\bL}{\mathbf{L}}
\newcommand{\bG}{\mathbf{G}}
\newcommand{\bW}{\mathbf{W}}
\newcommand{\bP}{\mathbf{P}}

\newcommand{\bC}{\mathbf{C}}
\newcommand{\bK}{\mathbf{K}}
\newcommand{\bM}{\mathbf{M}}

\newcommand{\bS}{\mathbf{S}}

\newcommand{\bY}{\mathbf{Y}}
\newcommand{\bZ}{\mathbf{Z}}

\newcommand{\bx}{\mathbf{x}}
\newcommand{\by}{\mathbf{y}}
\newcommand{\bv}{\mathbf{v}}

\newcommand{\bs}{\mathbf{s}}
\newcommand{\ba}{\mathbf{a}}
\newcommand{\bb}{\mathbf{b}}

\newcommand{\bc}{\mathbf{c}}
\newcommand{\bve}{\mathbf{e}}
\newcommand{\bu}{\mathbf{u}}

\newcommand{\bm}{\mathbf{m}}
\newcommand{\bg}{\mathbf{g}}

\newcommand{\bgamma}{\boldsymbol{\gamma}}

\newcommand{\blambda}{\boldsymbol{\lambda}}

\newcommand{\btheta}{\boldsymbol{\theta}}

\newcommand{\bpsi}{\boldsymbol{\psi}}

\newcommand{\balpha}{\boldsymbol{\alpha}}
\newcommand{\bbeta}{\boldsymbol{\beta}}

\newcommand{\0}{\mathbf{0}}

\def\R{\Bbb{R}}

\newtheorem{lem}{Lemma}
\newtheorem{rem}{Remark}

\newtheorem{exam}{Example}

\newcommand{\argmax}{\operatornamewithlimits{argmax}}

\begin{document}
\thispagestyle{empty} \baselineskip=28pt \vskip 5mm
\begin{center} {\Huge{\bf Multi-Level Restricted Maximum Likelihood
      Covariance Estimation and Kriging for Large Non-Gridded
      Spatial Datasets}}
\end{center}

\baselineskip=12pt \vskip 10mm

\begin{center}\large
Julio E. Castrill\'on-Cand\'as$^{1}$, Marc G.~Genton$^{2}$, and Rio
Yokota$^{3}$ \makeatletter{ \renewcommand*{\@makefnmark}{}
  \footnotetext{SRI Center for Uncertainty Quantification in
    Computational Science and Engineering$^{1}$; Computer, Electrical
    and Mathematical Sciences and Engineering$^{2}$, King Abdullah
    University of Science and Technology, Thuwal 23955-6900, Saudi
    Arabia; Tokyo Institute of Technology Global Scientific and
    Computing Center$^{3}$, 2-12-1 i7-2 O-okayama Meguro-ku, 152-8550,
    Tokyo, Japan.  E-mails: uvel@alum.mit.edu,
    marc.genton@kaust.edu.sa, rioyokota@gsic.titech.ac.jp}
  \makeatother}

\bigskip

{\it To appear in Spatial Statistics, (CC BY-NC-ND),
  doi:10.1016/j.spasta.2015.10.006.

}
\end{center}

\baselineskip=17pt \vskip 10mm \centerline{\today} \vskip 10mm

\begin{center}
{\large{\bf Abstract}}
\end{center}

\baselineskip=12pt

\begin{doublespace}
  We develop a multi-level restricted Gaussian maximum likelihood
  method for estimating the covariance function parameters and
  computing the best unbiased predictor. Our approach produces a new
  set of multi-level contrasts where the deterministic parameters of
  the model are filtered out thus enabling the estimation of the
  covariance parameters to be decoupled from the deterministic
  component. Moreover, the multi-level covariance matrix of the
  contrasts exhibit fast decay that is dependent on the smoothness of
  the covariance function.  Due to the fast decay of the multi-level
  covariance matrix coefficients only a small set is computed with a
  level dependent criterion.  We demonstrate our approach on problems
  of up to 512,000 observations with a Mat\'{e}rn covariance function
  and highly irregular placements of the observations. In addition,
  these problems are numerically unstable and hard to solve with
  traditional methods.

\end{doublespace}

\par\vfill\noindent {\bf KEY WORDS:} Fast Multipole Method; Hierarchical Basis; High
Performance Computing; Sparsification of Covariance Matrices

\par\medskip\noindent {\bf Short title}:
Multi-Level Restricted Maximum Likelihood and Kriging

\clearpage\pagebreak\newpage \pagenumbering{arabic}
\baselineskip=25.9pt

\section{Introduction}
\label{Introduction}

Consider the following model for a Gaussian spatial random field $Z$:
\begin{equation}
Z(\bs) = \bm(\bs)\T \bbeta+\epsilon(\bs), \qquad \bs \in \R^d,
\label{Introduction:noisemodel}
\end{equation}
where $\bm\in\R^p$ is a known function of the spatial location $\bs$,
$\bbeta\in\R^p$ is an unknown vector of coefficients, and
$\epsilon$ is a stationary mean zero Gaussian random field with
parametric covariance function
$C(\bs,\bs';\btheta)=\cov\{\epsilon(\bs),\epsilon(\bs')\}$
having an unknown vector $\btheta\in\R^w$ of parameters. We observe
the data vector $\bZ=(Z(\bs_1),\ldots,Z(\bs_n))\T$ at locations
$\mathbb{S}:=\{ \bs_{1},\dots,\bs_{n}\}$, where $\bs_{1} \neq
\bs_{2} \neq \bs_{3} \neq$ $\dots \neq \bs_{n-1} \neq \bs_{n}$, and
wish to: 1) estimate the unknown vectors $\bbeta$ and $\btheta$; and
2) predict $Z(\bs_0)$, where $\bs_0$ is a new spatial location. These
two tasks are particularly challenging when the sample size $n$ is
large.

To address the estimation part, let $\bC(\btheta)=\cov(\bZ,\bZ\T)\in
\R^{n \times n}$ be the covariance matrix of $\bZ$ and assume it is
nonsingular for all $\btheta\in\R^w$.  Define $\bM=\big( \bm(\bs_1)
\ldots \bm(\bs_n)\big)\T\in \R^{n\times p}$ and assume it is of full
rank, $p$. The model \eqref{Introduction:noisemodel} leads to the
vectorial formulation
\begin{equation}
{\bf Z} = {\bf M \bbeta}+{\boldsymbol \epsilon},
\label{Introduction:vectormodel}
\end{equation}
where $\boldsymbol \epsilon$ is a Gaussian random vector, ${\boldsymbol \epsilon}
\sim {\cal N}_{n}(\0,\bC(\btheta))$. Then the log-likelihood function
is
\begin{equation}
\ell(\bbeta,\btheta)=-\frac{n}{2}\log(2\pi)-\frac{1}{2}\log
\det\{\bC(\btheta)\}-\frac{1}{2}(\bZ-\bM\bbeta)\T\bC(\btheta)^{-1}(\bZ-\bM\bbeta),
\label{Introduction:loglikelihood}
\end{equation}
which can be profiled by generalized least squares with
\begin{equation}
\hat \bbeta(\btheta)=\{\bM\T \bC(\btheta)^{-1} \bM\}^{-1}\bM\T \bC(\btheta)^{-1}\bZ. \label{GLSbeta}
\end{equation}
A consequence of profiling is that the maximum likelihood estimator
(MLE) of $\btheta$ then tends to be biased.  A solution to this
problem is to use restricted maximum likelihood (REML) estimation
which consists in calculating the log-likelihood of $n-p$ linearly
independent contrasts, that is, linear combinations of observations
whose joint distribution does not depend on $\bbeta$, from the set
$\bY=\{\bI_n-\bM(\bM\T\bM)^{-1}\bM\T\}\bZ$.  In this paper, we propose
a new set of contrasts that lead to significant
computational benefits (with good accuracy) when computing the REML
estimator of $\btheta$ for large sample size~$n$.


To address the prediction part, consider the best unbiased
predictor $\hat Z(\bs_0)=\lambda_0+\blambda\T\bZ$ where
$\blambda=(\lambda_1,\ldots,\lambda_n)\T$. The unbiasedness constraint
implies $\lambda_0=0$ and $\bM\T\blambda=\bm(\bs_0)$.  The
minimization of the mean squared prediction error
E$[\{Z(\bs_0)-\blambda\T\bZ\}^2]$ under the constraint
$\bM\T\blambda=\bm(\bs_0)$ yields
\begin{equation}
\hat Z(\bs_0)=\bm(\bs_0)\T\hat \bbeta+\bc(\btheta)\T
\bC(\btheta)^{-1}(\bZ-\bM\hat \bbeta), \label{KrigBLUP}
\end{equation}
where $\bc(\btheta)=\cov\{\bZ,Z(\bs_0)\}\in \R^{n}$ and $\hat \bbeta$ is
defined in (\ref{GLSbeta}).  In this paper, we propose a new
transformation of the data vector $\bZ$ leading to a decoupled
multi-level description of the model \eqref{Introduction:noisemodel}
without any loss of structure. This multi-level representation leads
to significant computational benefits when computing the kriging
predictor $\hat Z(\bs_0)$ in (\ref{KrigBLUP}) for large sample size
$n$.

Previous work has been performed to maximize
\eqref{Introduction:loglikelihood}.  The classical technique is to
compute a Cholesky factorization of $\bC$. However, this requires
${\cal O}(n^2)$ memory and ${\cal O}(n^3)$ computational steps, thus
impractical for large scale problems.

Under special structures of the covariance matrix, i.e., fast decay of
the covariance function, a tapering technique can be used to sparsify
the covariance matrix and thus increase memory and computational
efficiency (\cite{Furrer2006,Kaufman2008}). These techniques
are good when applicable but tend to be restrictive. For a review of various
approaches to spatial statistics for large datasets, see \cite{Sun2012}.

Recently we have seen the advent of solving the optimization problem
\eqref{Introduction:loglikelihood} from a computational numerical
perspective.  \cite{Anitescu2012} developed a matrix-free approach for
computing the maximum of the log-likelihood
\eqref{Introduction:loglikelihood} based on a stochastic programming
reformulation.  This method relies on Monte Carlo approximation of the
derivative of the score function with respect to the covariance
parameters $\btheta$ to compute the maximization
\eqref{Introduction:loglikelihood}. The authors show promising results
for a grid geometry of the placement of the observations.  However for
a non-grid geometry the cost of computing the preconditioner becomes
${\cal O}(n^2)$ and it is not clear how many iterations for
convergence are needed as the geometry deviates from a
grid. \corb{Moreover, due to the slow convergence rate of the Monte
  Carlo method ($\eta^{-1/2}$ convergence rate where $\eta$ is the
  number of realizations) many samples might be required before a
  suitable estimate is obtained.} The previous work was extended in
\cite{Stein2013}.  Although the results are impressive (1,000,000 +
size problems), the approach is restricted to regular grid geometries
with partially occluded areas.

\cite{Stein2012} presented a difference filter preconditioning for
large covariance matrices not unlike our multi-level method. By
constructing a preconditioner based on the difference filter the
number of iterations of a Preconditioned Conjugate Gradient (PCG)
drops significantly. However, the authors can only construct a
preconditioner for irregularly placed observations in 1D and for a
regular grid in higher dimension. Moreover, the authors point out that
the restrictions on the spectral density of the random field $Z$
are strong.

In \cite{Stein2004} the authors proposed a REML method in combination
with an approximation of the likelihood. This approach uses a
truncation method to compute an approximation of the likelihood
function. It appears to be effective if the truncated terms have small
correlations. \corb{However, if the covariance function has a slow
  decay then we expect that this approximation will not be accurate
  unless a large neighborhood is incorporated.} Moreover, this paper
does not include an analysis of the error with respect to the
truncation.

In \cite{Sun2015} the authors proposed new unbiased estimating
equations based on score equation approximations. The inverse
covariance matrix is approximated with a sparse inverse Cholesky
decomposition.  \corb{As in \cite{Stein2004} the approximation is
  expected to be fast and accurate for locally correlated observations
  but will suffer from slow decay of the covariance function.}
Moreover, the results are limited to grid-like geometries.

In the next section we present the basic ideas behind our approach.
In Section \ref{MultiLevelREML} we show the construction of a
multi-level basis from the observations points. In Section
\ref{MultiLevelCovarianceMatrix} we describe how to efficiently construct
a multi-level covariance matrix that arises from the new basis. In
Section \ref{multilevelestimator} a multi-level estimator is proposed. In
Section \ref{multilevelkriging} the multi-level kriging approach is
described. In Section \ref{numericalstudy} hard to solve numerical
examples are provided and compared with traditional methods. In Section
\ref{multilevelconclusions} we give concluding remarks.  Proofs are
relegated to the Appendix A and a notation summary can be found in
Appendix B.  We also include computational and mathematical details in
the remarks. However, these may be skipped on a first reading except
for the more mathematically oriented reader.

\section{Multi-Level REML and Kriging Basic Approach}

We now present the main ideas of our proposal. Denote by ${\cal
  P}^{p}(\mathbb{S})$ the span of the columns of the design matrix
${\bf M}$. Let $\bL \in \R^{ p \times n} $ be an orthogonal projection
from $\R^n$ to ${\cal P}^{p}(\mathbb{S})$ and $\bW \in \R^{(n-p)
  \times n}$ be an orthogonal projection from $\R^n$ to ${\cal
  P}^{p}(\mathbb{S})^{\perp}$, the orthogonal complement of ${\cal
  P}^{p}(\mathbb{S})$.  Moreover we assume that the operator $\left[
\begin{array}{c}
\bW \\
\bL
\end{array}
\right ]$ is orthonormal.

By applying the operator ${\bf W}$ to
\eqref{Introduction:vectormodel} we obtain $\bZ_{W}={\bf W Z} = {\bf
  W} ({\bf M \bbeta}+ {\boldsymbol \epsilon}) = {\bf W{\boldsymbol
    \epsilon}}$. Our first observation is that the trend
contribution ${\bf M} \bbeta$ is filtered out from the data ${\bf
  Z}$. We can now formulate the estimation of the covariance
parameters $\btheta$ without the trend. The new log-likelihood
function becomes
\begin{equation}
\ell_{W}(\btheta)
=\corb{-\frac{n-p}{2}\log(2\pi)-\frac{1}{2}\log
\det\{\bC_{W}(\btheta)\}
-\frac{1}{2}\bZ_{W}\T\bC_{W}(\btheta)^{-1}\bZ_{W},}
\label{Introduction:multilevelloglikelihood}
\end{equation}
where $\bC_{W}(\btheta) = \bW \bC(\btheta) \bW \T$ and $\bZ_{W}\sim
{\cal N}_{n-p}(\0, \bW \bC(\btheta) \bW\T)$. As shown in Section
\ref{multilevelestimator} , to estimate the coefficients $\btheta$ it is
not necessary to compute $\ell_W(\btheta)$ but a multi-resolution
version.

A consequence of the filtering is that we obtain an unbiased
estimator.  Moreover, a further consequence is \corred{that if $\bv
  \neq 0$ then}
\begin{equation}
0 < \min_{\bv \in \mathbb{R}^{n}} \frac{\bv\T\bC(\btheta)\bv}{ \|\bv\|^2 } 
\leq 
\min_{\bv \in \mathbb{R}^{n} \backslash {\cal P}^{p}(\bbS)}
 \frac{\bv\T\bC(\btheta)\bv}{ \|\bv\|^2 } 
\leq \max_{\bv \in \mathbb{R}^{n} \backslash {\cal P}^{p}(\bbS)} \frac{\bv\T\bC(\btheta)\bv}
{ \|\bv\|^2 }  
\leq \max_{\bv \in \mathbb{R}^{n}} \frac{\bv\T\bC(\btheta)\bv}{ \|\bv\|^2 }.
\label{Introduction:eqn1}
\end{equation}
This implies that the condition number of $\bC_{W}(\btheta)$ is less
than or equal to the condition number of $\bC(\btheta)$. Thus
computing the inverse of $\bC_{W}(\btheta)$ will be in general more
stable than for $\bC(\btheta)$. In practice, computing the inverse of
$\bC_{W}(\btheta)$ will be much more stable than $\bC(\btheta)$ (See
the results in Tables \ref{SRBF:table1} and \ref{SRBF:table2}).
\corb{High condition number are very bad for numerical methods in
  general. In general any numerical method will suffer if the
  condition number is high.}  Finally, the uncertainties in the
parameter estimates obtained from
(\ref{Introduction:multilevelloglikelihood}) can be quantified using
the Godambe information matrix as described in Sect. 2 and Appendix B
of \cite{Stein2004}.


As shown in Section \ref{MultiLevelCovarianceMatrix}, for covariance
functions that are differentiable up to a degree $\tilde f + 1$
(except at the origin), such as the Mat\'{e}rn, our approach leads
to covariance matrices $\bC_{W}$ where most of the coefficients are
small and thus can be safely eliminated.  We construct a level
dependent criterion approach to determine which entries are computed
and the rest are set to zero. With this approach we can now construct
a sparse covariance matrix $\tilde{\bC}_{W}$ that is close to
$\bC_{W}$ in a matrix norm sense even if the observations are highly
correlated with distance. 

The sparsity of $\tilde{\bC}_{W}$ will depend on the following: i) a
positive integer $\tau$, which is a multi-level distance criterion;
ii) a positive integer $\tilde f$, which is the degree of the
multi-level basis and associated accuracy parameters $\tilde p$; and
iii) the smoothness of the covariance function. The accuracy of
$\tilde{\bC}_{W}$ will depend monotonically on these parameters, i.e.,
as we increase $\tau$ and $\tilde f$ (and respectively $\tilde p$) the
matrix $\tilde{\bC}_{W}$ will be closer to ${\bC}_{W}$ in a norm
sense. This is explained in detail in Section
\ref{MultiLevelCovarianceMatrix}.

The choice of the projectors $\bL$ and $\bW$ will determine how
efficiently each likelihood function
\eqref{Introduction:multilevelloglikelihood} evaluation is
solved. {\it Indeed, we desire the transformation to have the following
properties:}
i) {\bf Stability:} The matrices $\bL$ and $\bW$ have orthogonal
  rows and the stacked matrix $[\bL; \bW]$ is orthonormal;
ii) {\bf Fast computation:} The computational cost of applying the
  matrix $[\bL; \bW]$ to a vector is ${\cal
    O}(n(\log{ n})^{\xi})$ for some small integer $\xi$;
iii) {\bf Fast log determinant computation:} The computational cost
  of computing $\log \det\{\tilde{\bC}_{W}(\btheta)\}$ to be bounded 
   by ${\cal
    O}(n^{3/2})$ in 2D and ${\cal
    O}(n^{2})$ in 3D. We also want to restrict the memory storage to ${\cal
    O}(n (\log{n})^{\xi})$;
iv) {\bf Fast inversion:} The computational cost of computing
  $\bC_{W}(\btheta)^{-1} \bZ_{W}$ to a desired accuracy $\varepsilon$ 
is better than ${\cal
  O}(n^{2})$. Memory storage is also desirable to be restricted to
  ${\cal O}(n (\log{n})^{\xi})$;
v) {\bf Accuracy:} Determinant computation and inversion are
also required to be accurate. \corred{We achieve 
the properties i) - v) in this paper.}

In Section \ref{MultiLevelREML} we describe how to construct
multi-level matrices $\bL$ and $\bW$ that satisfy properties i) and
ii) for most practical observation location placements (random for
example). \corred{We apply $\bL$ and $\bW$ to construct the sparse
  multi-level covariance matrix $\tilde \bC_W(\btheta)$.} The
determinant of the multi-level sparse covariance matrix
\corred{$\tilde \bC_{W}(\btheta)$} and the term \corred{$\tilde
  \bC_{W}(\btheta)^{-1}\bZ_{W}$} are computed by exploiting an
accurate sparse \corred{Cholesky} representation \corred{of} $\bC_{W}$
(properties iii) and v) ). The term $\bC_{W}(\btheta)^{-1} \bZ_{W}$
can also be computed by applying a Preconditioned Conjugate Gradient
(PCG) to a desired accuracy (properties iv) and v) ).  In Section
\ref{multilevelkriging} the multi-level kriging method is described.
In Section \ref{numericalstudy} we demonstrate the efficiency of our
method for numerous covariances and irregularly placed observations.
We are able to solve the fast inversion for up to 512,000 observations
to a relative accuracy of $10^{-3}$ with respect to the {\it
  unpreconditioned} system. It is important to note that the achieved
accuracy of preconditioned system will not necessarily imply
accuracy of unpreconditioned system if the condition number of the
preconditioner is high.  Furthermore, we test our approach to estimate
the covariance parameters of problems of up to 128,000
observations. In addition, the accuracy of the kriging estimates are
tabulated for different size problems.






\section{Multi-Level Basis}
\label{MultiLevelREML}

In this section we establish the general structure of the Multi-Level
Basis (MB) that is used solve the estimation and prediction problem.
We refer the reader to \cite{Castrillon2013} for a detailed
description.  The MB can then be used to: (i) form the multi-level
REML function \eqref{Introduction:multilevelloglikelihood}; (ii)
sparsify the covariance matrix $\bC_{W}(\btheta)$; and (iii) improve
the conditioning over the covariance matrix $\bC(\btheta)$. But first,
we establish some notation and definitions:
\begin{itemize}

\item Let $\alpha:= (\alpha_{1},\dots,\alpha_{d}) \in \mathbb{Z}{^d}$,
  $|\alpha| := \alpha_{1}+\dots+\alpha_{d}$, $\bx : = [x_1,\dots,x_d]\T$
  and $D^\alpha_{\bx} := \frac{\partial^{\alpha_{1} + \dots +
      \alpha_{d}}} {\partial x_{1}^{\alpha_{1}} \dots \partial
    x_{d}^{\alpha_{d}}}$. For any $h \in \bbN_0$ (where $\mathbb{N}_0
  := \mathbb{N} \cup \{0\}$) let ${\cal Q}^d_h$ be the set of
  monomials $\{x_1^{\alpha_1} \dots x_d^{\alpha_d}\,\,\,|\,\,\,
  |\alpha| \leq h\}$. Furthermore, let $\bM_h$ be the design matrix
  with respect to all the monomials in ${\cal Q}^d_h$. The
  number of monomials of degree $h$ with dimension $d$ is
  $\begin{pmatrix} d + h \\ h \end{pmatrix}$.

\item We shall restrict the Gaussian spatial random field
  \eqref{Introduction:noisemodel} design matrix $\bM$ to $\bM_f$,
  where $f$ is the degree of the model. Thus $p$ will be equal to the
  number of monomials in ${\cal Q}^d_f$, which is $p
  := \begin{pmatrix} d + f \\ f \end{pmatrix}$.

\item Let $\tilde f \geq f$ be the degree of the multi-level basis and
  $\bM_{\tilde f}$ the associated design matrix.  The number of
  monomials in ${\cal Q}^d_{\tilde f}$ shall be referred as the
  accuracy parameter $\tilde p :=\begin{pmatrix} d + \tilde f
  \\ \tilde f \end{pmatrix}$.  These parameters are chosen by the
  user and are used to construct the multi-level basis.

\item Let $\bC(\btheta) : = \{ \phi(r_{i,j}; \btheta) \}$ where $\phi$
  is the covariance function, $r_{i,j} := \|\bs_i - \bs_j\|_2$ and
  $\bs_i,\bs_j \in \R^{d}$ for $i,j = 1, \dots, n$. Alternatively we
  refer to $\phi(\bx,\by; \btheta)$ as $\phi(r; \btheta)$, where
  $r:=\|\bx - \by\|_2$ and $\bx,\by \in \R^d$. Suppose ${\cal
    P}^{p}(\mathbb{S})$ is the span of the columns of the design
  matrix $\bM_f$.  We now assume that $\phi(r; \btheta)$ is a positive
  definite function and $C^{\tilde{f}+1}(\R)$ for all $r \in \R$
  except at the origin.

\item For any index $i,j \in \mathbb{N}_{0}$, $1 \leq i \leq n$, $1 \leq j
\leq n$, let $\bve_{i}[j] = \delta[i-j]$, where
$\delta[\cdot]$ is the discrete Kronecker delta function.

\end{itemize}

\begin{rem}
In practice instead of using the set of monomials ${\cal Q}^d_f$ we
use the set of Chebyshev polynomials of the first kind as these lead
to a more stable numerical scheme. However, for simplicity of the
presentation we keep it to monomials.
\end{rem}


The first step is to decompose the locations into a
series of multi-level cubes of dimension $d$. Without loss of
generality we assume that all the locations are contained in a unit
cube $B^{0}_{0}$ at level 0 and index 0. If the number of locations
inside $B^{0}_{0}$ is more than $p$ then equally subdivide $B^{0}_{0}$
into $2^{d}$ cubes ($B^{1}_{0}$, \dots, $B^{1}_{2^d - 1}$), where $d$ is the
number of dimensions. If the number of locations is $p$ or less then
stop, associate every location with the cube $B^{0}_{0}$ and denote
this as a leaf cube.  Otherwise, for each non-empty cube $B^{q}_{k}$ at
level $q = 1$ and index $k$ if the number of locations is more than
$p$ then subdivide, otherwise associate all the locations to
$B^{q}_{k}$ and denote this as a leaf cube.  This process is repeated for
all the subdivided cubes at levels $q = 2, \dots,$ until no
subdivisions are possible. The result is a tree structure with $0,
\ldots, t$ levels (See Algorithm~1 in \cite{Castrillon2013} for more
details).  We denote the leaf cubes as all the non-empty cubes that
contain at most $p$ locations, i.e., they will correspond to the leafs
of the tree structure.

\begin{rem}
For practical cases, $t$ increases proportionally to
$\log{n}$. If the inter location spacing collapses as $n^{-q}$ , where
$q$ is independent of $n$ , then $q\log{n}$ levels are needed, see
Section 4 in \cite{Beatson1997} for details.
\end{rem}

Suppose that there is a one-to-one mapping between the set of unit
vectors ${\cal E}:=\{\bve_{1},\dots,\bve_{n}\}$, which we denote as
leaf unit vectors, and the set of locations $\{
\bs_{1},\dots,\bs_{n}\}$, i.e. $\bs_{i} \longleftrightarrow \bve_{i}$
for all $i = 1, \dots, n$. It is clear that the space of
$\{\bve_{1},\dots,\bve_{n}\}$ is $\bbR^{n}$. The next step is to replace
${\cal E}$ with a new basis of $\R^{n}$ that is multi-level,
orthonormal and gives us the desired properties i) and ii) from Section
\ref{Introduction}. In \cite{Castrillon2013} the reader can find full
details on such a construction. However, for the sake of clarity for
the rest of the paper we associate the multi-level domain decomposition
to the multi-level basis:

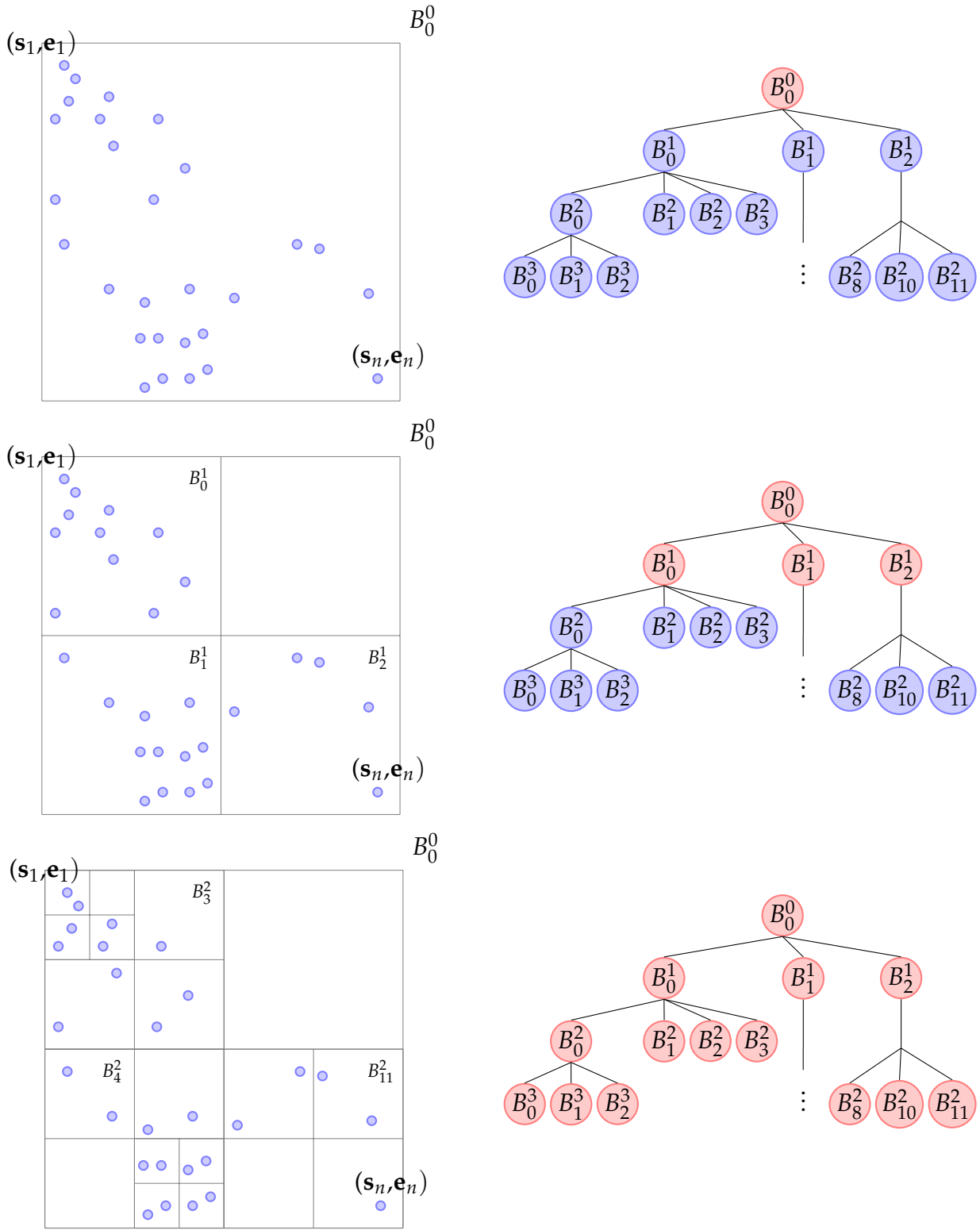
\begin{figure}

\vspace{-5cm}

\begin{tabular}{c c}

\begin{tikzpicture}[scale=.75] 
  \begin{scope} 
    [place/.style={circle,draw=blue!50,fill=blue!20,thick,
      inner sep=0pt,minimum size=1.5mm}]
    \draw[step=8,gray,very thin] (0, 0) grid (8, 8);

    \node[anchor=center] at (0, -4.5) {$$};
    \node[anchor=center] at (0,   10) {$$};

    \node at (0.5,7.5) [place] {};
    \node at (0,8.0) [] {($\bs_1$,$\bve_1)$};

    \node at (0.75,7.2) [place] {};
    \node at (0.3,6.3) [place] {};
    \node at (0.6,6.7) [place] {};
    \node at (1.3,6.3) [place] {};
    \node at (1.5,6.8) [place] {};

    \node at (2.5,4.5) [place] {};
    \node at (0.3,4.5) [place] {};
    \node at (1.6,5.7) [place] {};
    \node at (2.6,6.3) [place] {};
    \node at (3.2,5.2) [place] {};

    \node at (0.5,3.5) [place] {};
    \node at (1.5,2.5) [place] {};
    \node at (2.3,2.2) [place] {};
    \node at (3.3,2.5) [place] {};

    \node at (2.3,0.3) [place] {};
    \node at (2.7,0.5) [place] {};
    \node at (2.2,1.4) [place] {};
    \node at (2.6,1.4) [place] {};
    \node at (3.3,0.5) [place] {};
    \node at (3.7,0.7) [place] {};
    \node at (3.2,1.3) [place] {};
    \node at (3.6,1.5) [place] {};

    \node at (4.3,2.3) [place] {}; \node at (5.7,3.5) [place] {};
    \node at (6.2,3.4) [place] {}; \node at (7.3,2.4) [place] {};
    \node at (7.5,0.5) [place] {}; \node at (7.75,1) [] {($\bs_n$,$\bve_n)$};
    \node at (8.5,8.5) [] {$B^{0}_0$};
  \end{scope}

\end{tikzpicture}
\vspace{-8cm}
& 

\begin{tikzpicture}[scale=1]

    \node[anchor=center] at (0, -4.5) {$$};
    \node[anchor=center] at (0,   10) {$$};

\begin{scope}[xshift=5cm, yshift=4cm,
place/.style={circle,draw=blue!50,fill=blue!20,thick,
      inner sep=0pt,minimum size=1.5mm},
placer/.style={circle,draw=red!50,fill=red!20,thick,
      inner sep=0pt,minimum size=1.5mm},
]
\Tree [.\node[placer]{$B^{0}_{0}$}; 
             [.\node[place]{$B^{1}_{0}$};
                    [.\node[place]{$B^{2}_{0}$}; 
                            \node[place]{$B^{3}_{0}$}; 
                            \node[place]{$B^{3}_{1}$}; 
                            \node[place]{$B^{3}_{2}$}; 
                    ] 
                    [.\node[place]{$B^{2}_{1}$}; ] 
                    [.\node[place]{$B^{2}_{2}$}; ] 
                    [.\node[place]{$B^{2}_{3}$}; ] 
             ]                                         
             [.\node[place]{$B^{1}_{1}$}; 
                    [      [. $\vdots$ ] ]
             ]   
             [.\node[place]{$B^{1}_{2}$};
                    [      [.\node[place]{$B^{2}_{8}$}; ]
                           [.\node[place]{$B^{2}_{10}$}; ] 
                           [.\node[place]{$B^{2}_{11}$}; ]
                                          ]]] 
\end{scope}
\end{tikzpicture} 

\\

\begin{tikzpicture}[scale=.75] 
  \begin{scope} 
    [place/.style={circle,draw=blue!50,fill=blue!20,thick,
      inner sep=0pt,minimum size=1.5mm}]
    \draw[step=4,gray,very thin] (0, 0) grid (8, 8);

    \node[anchor=center] at (0, -4.5) {$$};
    \node[anchor=center] at (0,   10) {$$};

    \node at (0.5,7.5) [place] {};
    \node at (0,8.0) [] {($\bs_1$,$\bve_1)$};

    \node at (0.75,7.2) [place] {};
    \node at (0.3,6.3) [place] {};
    \node at (0.6,6.7) [place] {};
    \node at (1.3,6.3) [place] {};
    \node at (1.5,6.8) [place] {};

    \node at (2.5,4.5) [place] {};
    \node at (0.3,4.5) [place] {};
    \node at (1.6,5.7) [place] {};
    \node at (2.6,6.3) [place] {};
    \node at (3.2,5.2) [place] {};

    \node at (0.5,3.5) [place] {};
    \node at (1.5,2.5) [place] {};
    \node at (2.3,2.2) [place] {};
    \node at (3.3,2.5) [place] {};

    \node at (2.3,0.3) [place] {};
    \node at (2.7,0.5) [place] {};
    \node at (2.2,1.4) [place] {};
    \node at (2.6,1.4) [place] {};
    \node at (3.3,0.5) [place] {};
    \node at (3.7,0.7) [place] {};
    \node at (3.2,1.3) [place] {};
    \node at (3.6,1.5) [place] {};

    \node at (4.3,2.3) [place] {}; \node at (5.7,3.5) [place] {};
    \node at (6.2,3.4) [place] {}; \node at (7.3,2.4) [place] {};
    \node at (7.5,0.5) [place] {}; \node at (7.75,1) [] {($\bs_n$,$\bve_n)$};
    \node at (8.5,8.5) [] {$B^{0}_0$};
    \node at (3.5,3.5) [scale=0.75] {$B^1_1$};
    \node at (7.5,3.5) [scale=0.75] {$B^1_2$};
    \node at (3.5,7.5) [scale=0.75] {$B^1_0$};

 \end{scope}
\end{tikzpicture} 
\vspace{-8cm}

& 

\begin{tikzpicture}[scale=1]

    \node[anchor=center] at (0, -4.5) {$$};
    \node[anchor=center] at (0,   10) {$$};

\begin{scope}[xshift=5cm, yshift=4cm,
place/.style={circle,draw=blue!50,fill=blue!20,thick,
      inner sep=0pt,minimum size=1.5mm},
placer/.style={circle,draw=red!50,fill=red!20,thick,
      inner sep=0pt,minimum size=1.5mm}
]
\Tree [.\node[placer]{$B^{0}_{0}$}; 
             [.\node[placer]{$B^{1}_{0}$};
                    [.\node[place]{$B^{2}_{0}$}; 
                            \node[place]{$B^{3}_{0}$}; 
                            \node[place]{$B^{3}_{1}$}; 
                            \node[place]{$B^{3}_{2}$}; 
                    ] 
                    [.\node[place]{$B^{2}_{1}$}; ] 
                    [.\node[place]{$B^{2}_{2}$}; ] 
                    [.\node[place]{$B^{2}_{3}$}; ] 
             ]                                         
             [.\node[placer]{$B^{1}_{1}$}; 
                    [      [. $\vdots$ ] ]
             ]   
             [.\node[placer]{$B^{1}_{2}$};
                    [      [.\node[place]{$B^{2}_{8}$}; ]
                           [.\node[place]{$B^{2}_{10}$}; ] 
                           [.\node[place]{$B^{2}_{11}$}; ]
                                          ]]] 
\end{scope}
\end{tikzpicture} 

\\

\begin{tikzpicture}[scale=.75] 
  \begin{scope} 
    [place/.style={circle,draw=blue!50,fill=blue!20,thick,
      inner sep=0pt,minimum size=1.5mm}]

    \draw[step=4,gray,very thin] (0, 0) grid (8, 8);
    \draw[step=2,gray,very thin] (0, 0) grid (4, 4);
    \draw[step=2,gray,very thin] (0, 4) grid (4, 8);
    \draw[step=2,gray,very thin] (4, 0) grid (8, 4);

    \draw[step=1,gray,very thin] (0, 6) grid (2, 8);
    \draw[step=1,gray,very thin] (2, 0) grid (4, 2);

    \node[anchor=center] at (0, -4.5) {$$};
    \node[anchor=center] at (0,   10) {$$};

    \node at (0.5,7.5) [place] {};
    \node at (0,8.0) [] {($\bs_1$,$\bve_1)$};

    \node at (0.75,7.2) [place] {};
    \node at (0.3,6.3) [place] {};
    \node at (0.6,6.7) [place] {};
    \node at (1.3,6.3) [place] {};
    \node at (1.5,6.8) [place] {};

    \node at (2.5,4.5) [place] {};
    \node at (0.3,4.5) [place] {};
    \node at (1.6,5.7) [place] {};
    \node at (2.6,6.3) [place] {};
    \node at (3.2,5.2) [place] {};

    \node at (0.5,3.5) [place] {};
    \node at (1.5,2.5) [place] {};
    \node at (2.3,2.2) [place] {};
    \node at (3.3,2.5) [place] {};

    \node at (2.3,0.3) [place] {};
    \node at (2.7,0.5) [place] {};
    \node at (2.2,1.4) [place] {};
    \node at (2.6,1.4) [place] {};
    \node at (3.3,0.5) [place] {};
    \node at (3.7,0.7) [place] {};
    \node at (3.2,1.3) [place] {};
    \node at (3.6,1.5) [place] {};

    \node at (4.3,2.3) [place] {}; \node at (5.7,3.5) [place] {};
    \node at (6.2,3.4) [place] {}; \node at (7.3,2.4) [place] {};
    \node at (7.5,0.5) [place] {}; \node at (7.75,1) [] {($\bs_n$,$\bve_n)$};
    \node at (8.5,8.5) [] {$B^{0}_0$};
    \node at (3.5,7.5) [scale=0.75] {$B^2_3$};
    \node at (7.5,3.5) [scale=0.75] {$B^2_{11}$};
    \node at (1.5,3.5) [scale=0.75] {$B^2_4$};

 \end{scope}
\end{tikzpicture} 
&

\begin{tikzpicture}[scale=1]

    \node[anchor=center] at (0, -4.5) {$$};
    \node[anchor=center] at (0,   10) {$$};

\begin{scope}[xshift=5cm, yshift=4cm,
placer/.style={circle,draw=red!50,fill=red!20,thick,
      inner sep=0pt,minimum size=1.5mm}]
\Tree [.\node[placer]{$B^{0}_{0}$}; 
             [.\node[placer]{$B^{1}_{0}$};
                    [.\node[placer]{$B^{2}_{0}$}; 
                            \node[placer]{$B^{3}_{0}$}; 
                            \node[placer]{$B^{3}_{1}$}; 
                            \node[placer]{$B^{3}_{2}$}; 
                    ] 
                    [.\node[placer]{$B^{2}_{1}$}; ] 
                    [.\node[placer]{$B^{2}_{2}$}; ] 
                    [.\node[placer]{$B^{2}_{3}$}; ] 
             ]                                         
             [.\node[placer]{$B^{1}_{1}$}; 
                    [      [. $\vdots$ ] ]
             ]   
             [.\node[placer]{$B^{1}_{2}$};
                    [      [.\node[placer]{$B^{2}_{8}$}; ]
                           [.\node[placer]{$B^{2}_{10}$}; ] 
                           [.\node[placer]{$B^{2}_{11}$}; ]
                                          ]]] 
\end{scope}
\end{tikzpicture} 

\\

\end{tabular}

\vspace{-3cm}

\caption{Multi-level domain decomposition of the location of
  observations for $d = p =2$. All the observation locations $\bs_i$
  are assumed to be contained in the unit cube $B^{0}_{0}$ (colored red
  node in tree). If the number of observations are greater than $p =
  2$ we subdivide into equal boxes. At the end we obtain a multi-level
  decomposition of the observation points.}
\label{MLRLE:fig1}
\end{figure}

\begin{itemize}
\item 
For each non empty cube $B^i_k$, for $i = 0, \dots, t$, associate a
series of multi-level basis vectors $\{ \bpsi^{i,k}_{\tilde k_1},
\bpsi^{i,k}_{\tilde k_2}, \dots \}$ that have the following property:
\begin{equation}
\bg^T  \bpsi^{i,k}_{\tilde k_j} = \sum_{a=1}^{n} \bg[a] \bpsi^{i,k}_{\tilde k_j}[a] = 0,
\label{hbconstruction:eqn2}
\end{equation}
for $j = 0,\dots$ and for all the vectors $\bg$ that are columns of
$\bM_{\tilde f}$. Furthermore, let $\bW^{i,k}$ be a matrix
$[ \bpsi^{i,k}_{\tilde k_1}, \bpsi^{i,k}_{\tilde k_2}, \dots ]$.

\item For $i = 0, \dots, t$ let $\bW_{i} := [\bW^{i,0}, \bW^{i,1},
  \dots]$

\item If $\tilde p > p$ we will have an extra $\tilde p - p$
  vectors corresponding to a level $-1$ for the initial cube
  $B^{0}_{0}$. Now, associate $\tilde p - p$ multi-level vectors $\{
  \bpsi^{-1,0}_{0},\bpsi^{-1,0}_{1}, \dots, \bpsi^{-1,0}_{\tilde p - p}
  \}$ that have the following property:
\begin{equation}
\bg\T \bpsi^{-1,0}_{j} =
\sum_{a=1}^{n} \bg[a] \bpsi^{-1,0}_{j}[a] = 0,
\label{hbconstruction:eqn3}
\end{equation}
for $j = 1,\dots, \tilde p - p$ and for all the vectors $\bg$ that are
columns of $\bM_{\tilde f}$. Similarly as above, let $\bW_{-1}:= [
  \bpsi^{-1,0}_{0},\bpsi^{-1,0}_{1}, \dots, \bpsi^{-1,0}_{\tilde p - p}
]$.

\item In total we will have $n - p$ multi-level vectors
and the transform matrix $\bW \in \R^{(p - n) \times n}$ is built as
$\bW : = [\bW_t, \dots, \bW_0, \bW_{-1}]^T$.

\item Now, it is clear that $\bW \bg = {\bf 0}$ for any $\bg \in
  \bM_{f}$. To complete the basis to span $\R^{n}$ we need $p$ more
  orthonormal vectors. In \cite{Castrillon2013} it is shown how to
  compute such a basis and stack the vectors as rows in the matrix
  $\bL \in \R^{p \times n}$.

\end{itemize}

With the construction of $\bW$ and $\bL$ we will have the following
properties: a) the matrix $ \bP := \left[
\begin{array}{c}
\bW \\
\bL
\end{array}
\right
]$ 
is orthonormal, i.e., $\bP\bP\T= \bI_n$;
b) any vector $\bv \in \R^{n}$ can be written as $\bv = \bL\T\bv_{L} +
  \bW\T\bv_{W}$ where $\bv_{L} \in \R^{p} $ and $\bv_{W} \in \R^{n-p}$ are
  unique; c) the matrix $\bW$ contains at most ${\cal O}(nt)$ non-zero
  entries and $\bL$ contains at most ${\cal O}(np)$ non-zero
  entries. This implies that for any vector $\bv \in \R^{n}$ the
  computational cost of applying $\bW \bv$ is at most ${\cal
    O}(nt)$ and $\bL \bv$ is at most ${\cal O}(np)$.
 
\section{Multi-Level Covariance Matrix}
\label{MultiLevelCovarianceMatrix}

In this section we show how we can use the matrix $\bW$ to produce a
highly sparse representation of $\bC_W(\btheta)$ with a
level-dependent tapering technique. 

\begin{figure}[h!] 
\psfrag{a}[l]{$\bC^{t,t}_{W}$} 
\psfrag{b}[c]{$\dots$}
  \psfrag{c}[c]{\hspace{1mm} {$\bC^{t,i}_{W}$} } 
  \psfrag{d}[c]{ \hspace{5mm}  $\vdots$}
  \psfrag{e}[l]{$\ddots$} 
  \psfrag{f}[c]{$\vdots$}
  \psfrag{g}[c]{\hspace{5mm} $\bC^{-1,t}_{W}$} 
  \psfrag{h}[c]{$\dots$}
  \psfrag{i}[c]{\small $\bC^{i,i}_{W}$}
\begin{center}
\includegraphics[width=2.7in,height=2.10in,trim=0 0 110 0,clip]{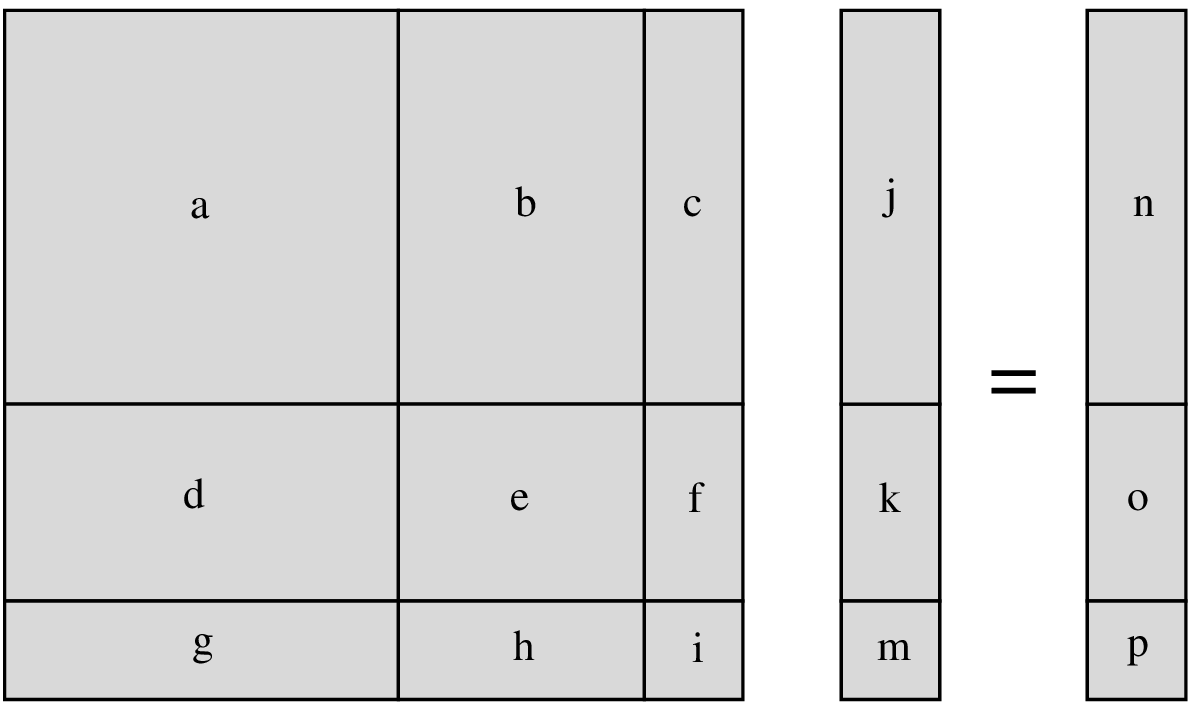}
\end{center}
\caption{Organization of the multi-level covariance matrix
  $\bC_W(\btheta)$. For this figure $i = -1$.}
\label{multilevelcov:fig1}
\end{figure}

With the MB we can transform the observation data vector $\bZ$ by
applying the matrix $\bW$. This leads to the multi-level
log-likelihood function
\eqref{Introduction:multilevelloglikelihood}. The covariance matrix
$\bC(\btheta)$ is now transformed into $\bC_{W}(\btheta)$ with the
structure shown in Figure \ref{multilevelcov:fig1} where each of the
blocks $\bC^{i,j}_W = \bW_i \bC(\btheta) \bW_j \T$ for all $i,j = 0,
\dots, t$. This implies that 
the entries of the matrix $\bC^{i,j}_W$ are formed from all the
interactions of the MB vectors between level $i$ and $j$. Thus for any
$\bpsi^{i,k}_{\tilde{l}}$ and $\bpsi^{j,l}_{\tilde{k}}$ vectors there
is a unique entry of $\bC^{i,j}_W$ of the form $
(\bpsi^{i,k}_{\tilde{k}})\T \bC(\btheta) \bpsi^{j,l}_{\tilde{l}}$.
The blocks $\bC_W^{i,j}$, where $i=-1$ or $j=-1$, correspond to the
case where the accuracy term $\tilde{p} > p$.

The following lemma relates the covariance function $\phi$, the degree
$\tilde{f}$ (corresponding to the accuracy parameter $\tilde{p}$) of
the design matrix $\bM_{\tilde{f}}$ to the decay of the entries of the
matrix $\bC_W(\btheta)$.
\begin{lem} Let $B_{\ba}$ be the smallest ball in ${\mathbb R}^{d}$
with radii $r_{\ba}$ centered around the midpoint $\ba \in {\mathbb
  R}^{d}$ of the cube $B^{i}_{l}$ such that $B^{i}_{l} \subset
B_{\ba}$. Similarly, let $B_{\bb}$ be the smallest ball in ${\mathbb
  R}^{d}$ with radii $r_{\bb} \in {\mathbb R}^{d}$ centered around the
midpoint $\bb$ of the cube $B^{j}_{k}$ such that $B^{j}_{k} \subset
B_{\bb}$.  Now, since $\bpsi^{i,l}_{\tilde{l}}$ and
$\bpsi^{j,k}_{\tilde{k}}$ satisfy the moment
orthogonality condition from equations \eqref{hbconstruction:eqn2} 
and \eqref{hbconstruction:eqn3} for
all $\bg \in {\cal P}^{\tilde{p}}(\bbS)$ then the following bound holds:
\begin{equation}
|(\bpsi^{i,k}_{\tilde{k}})\T \bC(\btheta) \bpsi^{j,l}_{\tilde{l}} |
\leq
\sum_{|\alpha|=\tilde{f}+1} \sum_{|\beta|=\tilde{f}+1}
\frac{r_{\ba}^{\alpha}}{\alpha!} \frac{r_{\bb}^{\beta}}{\beta!}
\sup_{\bx \in B_{\ba}, \by \in B_{\bb}}
 |D^{\alpha}_{\bx} D^{\beta}_{\by} \phi(\bx,\by;\btheta)|,
\label{multilevelcov:eqn1}
\end{equation}
\label{multilevelcov:lemma1}
for $i,j = 0,\dots,t$.
\end{lem}
From Lemma \ref{multilevelcov:lemma1} we observe that the decay of the
entries of $\bC_W(\btheta)$ is dependent on the magnitude of the
derivatives of the covariance function $\phi(r;\btheta)$, the size of
$B_{\ba}$ and $B_{\bb}$ and the degree of ${\cal Q}^d_{\tilde{f}}$. Thus if
$\phi(r;\btheta)$ is smooth on $B_{\ba}$ and $B_{\bb}$ the entries of
$\bC_W(\btheta)$ will be small.

\begin{exam}
In Figure \ref{multilevelcov:fig2} we show a comparison between (a)
the covariance matrix $\bC(\btheta)$ and (b) the multi-level
covariance matrix $\bC_W(\btheta)$ for the following example: 1)
$\phi(r;\btheta) : = \exp(-r)$ and $d = 3$.  2) The observation
locations ($n = 8000$) are sampled from a uniform distribution on the
unit cube $[0,1]^3$. The actual values of the observations are not
necessary for this example.  3) $f = 3$ (leading to $p = 20$
monomials).  4) We sort the $x_1$ direction location from 0 to 1. This
is done for visualization reasons so that we may observe the decay
in the matrix $\bC(\btheta)$.
\begin{figure}[h!]
\begin{center}
\begin{tabular}{cc}
   \includegraphics[ width=3.0in, height=3in
   ]{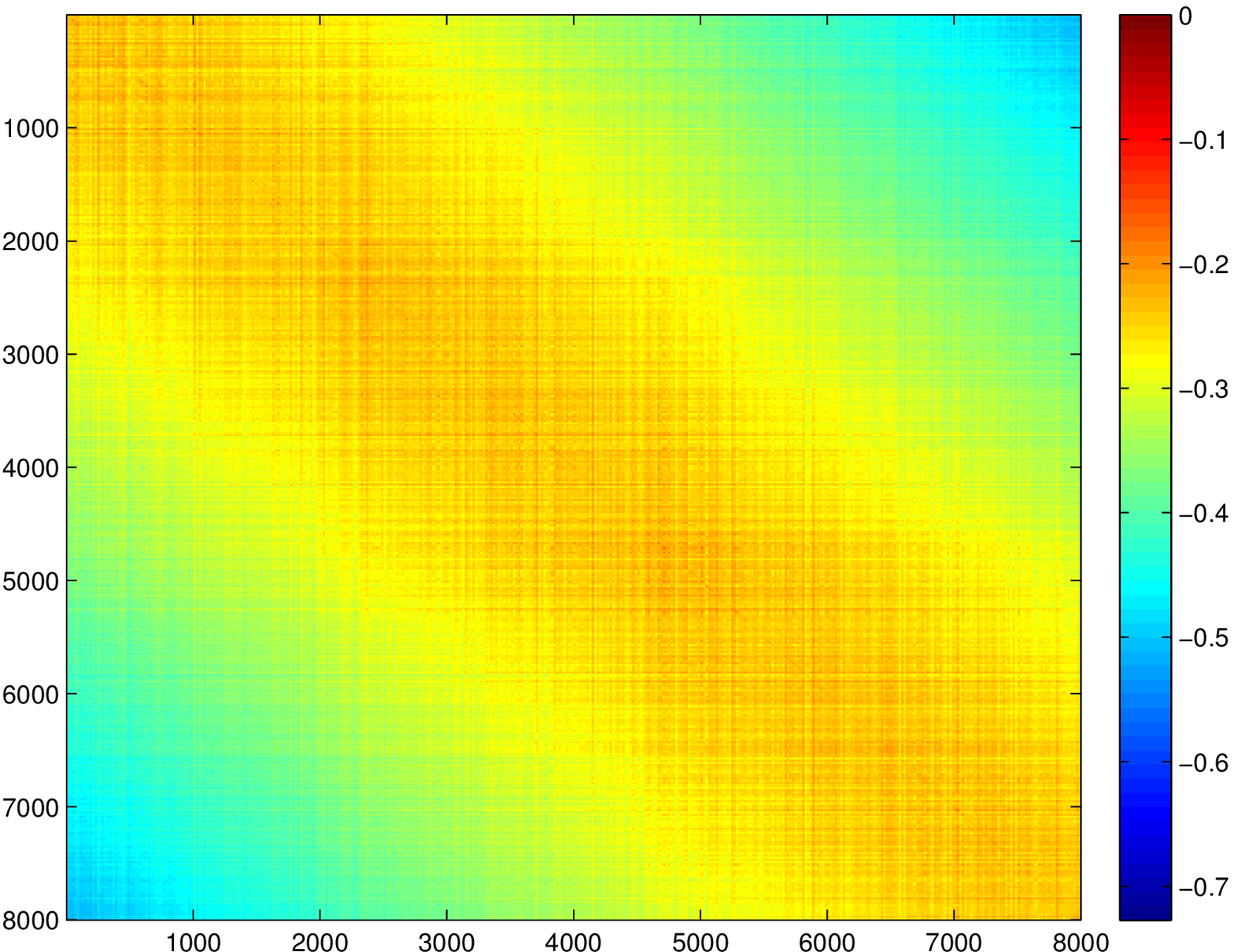} & \includegraphics[ width=3.0in,
     height=3in ]{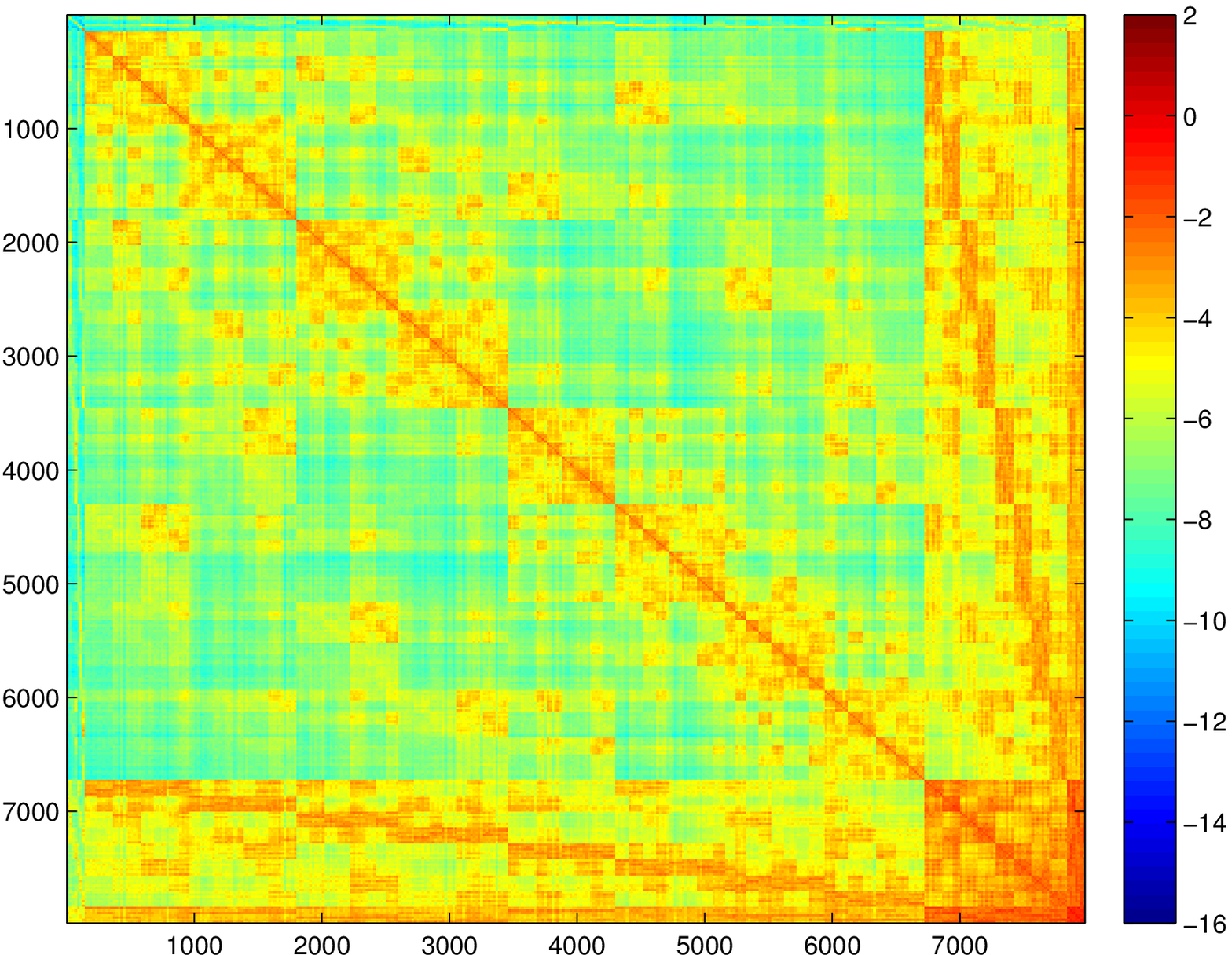} \\ (a)
   $\log_{10}{\mbox{abs}(\bC(\btheta))}$ & (b) $\log_{10}{\mbox{abs}(\bC_W(\btheta))}$ \\
\end{tabular} 
\end{center}
  \caption{Covariance matrix comparison between covariance matrices
    (a) $\log_{10}{\mbox{abs}(\bC(\btheta))}$ and (b)
    $\log_{10}{\mbox{abs}(\bC_W(\btheta))}$ for the exponential covariance
    function $\phi(r) = \exp(-r)$ with $n = 8000$, $p = 20$ and $d =
    3$.}
\label{multilevelcov:fig2}
\end{figure}
Notice that the decay of $\bC(\btheta)$ is dependent on the covariance
function $\phi(r;\btheta)$. It is clear that for this case a tapering
technique would not be very effective as most of the entries are
comparable in magnitude. In contrast a few of the entries of
$\bC_W(\btheta)$ with high magnitude are concentrated around
particular regions while most of the entries have very small
magnitudes making a hierarchical tapering technique to sparsify the
matrix a viable option.
\label{multilevelcov:example1}
\end{exam}

To produce a sparse matrix from $\bC_W(\btheta)$ we execute the
following multi-level tapering technique: 

\begin{itemize}
\item For all cubes $B^i_k$ at level $i$ let $L^{i,0}_k := B^i_k$ and
  $L^{i,j}_k := L^{i,j-1}_k \cup \{$union of all cubes at level $i$
  that share a face or corner with $L^{i,j-1}_k \}$ for $j =
  0,1,\dots$. A construction example is shown in Figure
  \ref{multilevelcov:fig3} for level $i$. Now, perform this
  construction for $i = 0,\dots,t$.

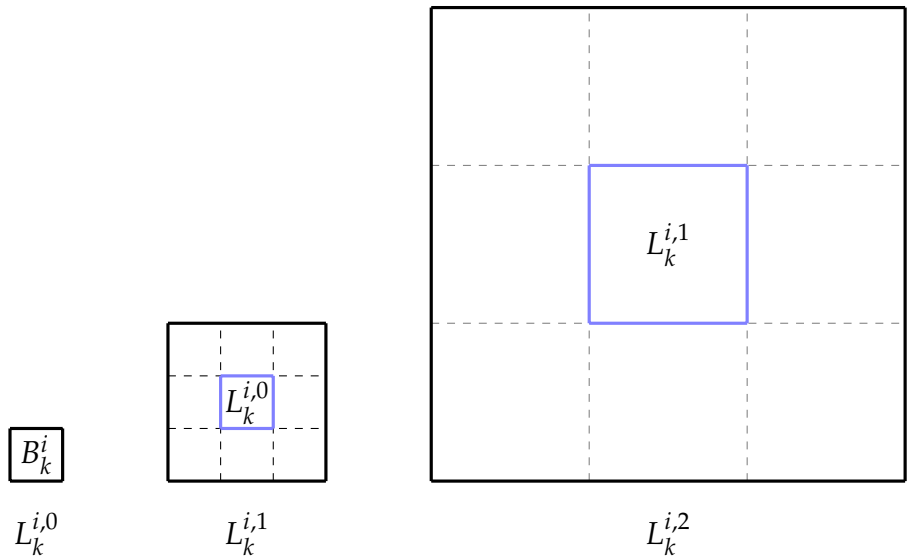
\begin{figure}[!h]
\begin{center}
\begin{tikzpicture}[scale=0.7] 
 \begin{scope} 
   [place/.style={circle,draw=blue!50,fill=blue!20,thick,
       inner sep=0pt,minimum size=0.5mm}]

    \draw[step=1, very thick] (1, 0) grid (2, 1);
    \node at (1.5,0.5) [] {$B^i_k$};
    \node at (1.5,-1) [] {$L^{i,0}_k$};

    \draw[step=1, dashed] (4, 0) grid (7, 3);
    \draw[step=1,blue!50, very thick ] (5, 1) grid (6, 2);

    \draw[very thick] (4, 0) -- (7, 0);
    \draw[very thick] (7, 0) -- (7, 3);
    \draw[very thick] (7, 3) -- (4, 3);
    \draw[very thick] (4, 3) -- (4, 0);

    \node at (5.5,1.5) [] {$L^{i,0}_k$};
    \node at (5.5,-1) [] {$L^{i,1}_k$};
    
    \draw[step=3,gray, dashed] (9, 0) grid (18, 9);
    \draw[step=3,blue!50, very thick] (12, 3) grid (15, 6);
    
    \draw[very thick] (9,  0) -- (18, 0);
    \draw[very thick] (18, 0) -- (18, 9);
    \draw[very thick] (18, 9) -- (9, 9);
    \draw[very thick] (9, 9) -- (9, 0);

    \node at (13.5,4.5) [] {$L^{i,1}_k$};
    \node at (13.5,-1) [] {$L^{i,2}_k$};
   
\end{scope}
\end{tikzpicture}
\end{center}

\caption{Construction of expanded cubes $L^{i,\tau}_{k}$, $\tau = 0,
  1, \dots$ from initial cube $B^{i}_{k}$.}
\label{multilevelcov:fig3}
\end{figure}

\item Set a user given constant $\tau \in \mathbb{N}_0$

\item The entry of $\bC_W(\btheta)$ corresponding to
  $(\bpsi^{i,k}_{\tilde k})\T \bC(\btheta) \bpsi^{j,l}_{\tilde l}$,
  for $i,j = 0,\dots,\tau$ is computed if the following level
  dependent criterion is true: If $\mbox{($j \geq i$ and $B^{j}_{l}
    \subset L^{i,\tau}_{k}$) or ($j < i$ and $B^{i}_{k} \subset
    L^{j,\tau}_{l}$)}$ is true for the given $\tau \in \mathbb{N}_0$
  then compute the entry $(\bpsi^{i,k}_{\tilde k})\T \bC(\btheta)
  \bpsi^{j,l}_{\tilde l}$.

\item For the case that $i = -1$ or $j = -1$ the entry
  corresponding to $(\bpsi^{i,k}_{\tilde k})\T \bC(\btheta)
  \bpsi^{j,l}_{\tilde l}$ is always computed.




\end{itemize}

\begin{figure}[!h]
\begin{center}
\begin{tikzpicture}[scale=0.7] 
 \begin{scope} 
   [place/.style={circle,draw=blue!50,fill=blue!20,thick,
       inner sep=0pt,minimum size=0.5mm}]




    
    \draw[step=1, gray,  dashed] (9, 0) grid (18, 9);

    \draw[step=1, blue!50, very thick] (16, 6) grid (17, 7);
    \node at (16.5,6.5) [ ] {$B^{j}_{l_1}$};

    \draw[step=1, blue!50, very thick] (11, 7) grid (12, 8);
    \node at (11.5,7.5) [ ] {$B^{j}_{l_2}$};

    \draw[step=1, blue!50, very thick] (10, 1) grid (11, 2);
    \node at (10.5,1.5) [ ] {$B^{j}_{l_3}$};

    \draw[step=1, blue!50, very thick] (14, 2) grid (15, 3);
    \node at (14.5,2.5) [ ] {$B^{j}_{l_4}$};

    \draw[step=1, blue!50, very thick] (16, 4) grid (17, 5);
    \node at (16.5,4.5) [ ] {$B^{j}_{l_5}$};

    \draw[step=1, very thick] (13, 4) grid (14, 5);
    \node at (13.5,4.5) [ ] {$B^{i}_k$};

    \draw[very thick] (9,  0) -- (18, 0);
    \draw[very thick] (18, 0) -- (18, 9);
    \draw[very thick] (18, 9) -- (9, 9);
    \draw[very thick] (9, 9) -- (9, 0);

    \node at (13.5,-1) [] {$B^j_{l_1}, \dots, B^j_{l_5} \subset
      L^{i,\tau}_k$; $\tau = 2$.};

    \node at (9.5,8.5) [] {$L^{i,\tau}_k$};
   
\end{scope}
\end{tikzpicture}
\end{center}

\caption{Example of finding all the boxes $B^j_{l}$ that are
contained in $L^{i,\tau}_k$ for $\tau = 2$.}
\label{multilevelcov:fig5}
\end{figure}
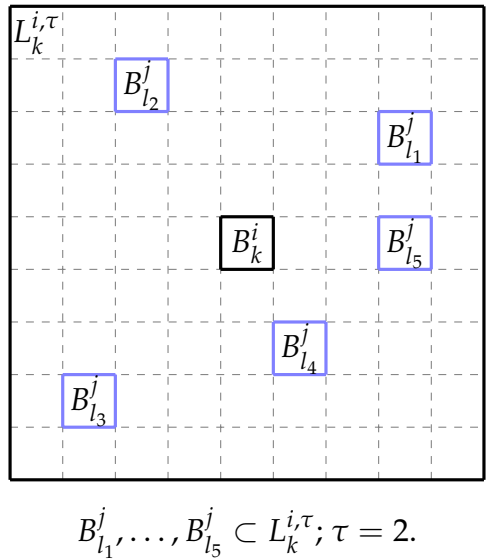

From this distance criterion we can apriori determine which entries of
the sparse matrix $\tilde \bC_W(\btheta)$ are to be computed. For any
given row of the matrix $\tilde \bC_W(\btheta)$ corresponding to level
$i$ and index $k$ construct the expanded cube $L^{i,\tau}_{k}$. Now,
for $j = i, \dots, t$ find all the cubes $B^j_l$ with the
corresponding index $l$ that are contained in $L^{i,\tau}_{k}$ (See
Figure \ref{multilevelcov:fig5}). For $j = 0, \dots, i - 1$ find all
the extended cubes $L^{j,\tau}_{l}$ such that $B^i_k \subset
L^{j,\tau}_{l}$. For $i,j = 0,\dots,t$ this action can be performed
efficiently by using the tree shown in Figure \ref{MLRLE:fig1}. 

With this criterion we can produce a highly sparse matrix
$\tilde{\bC}_{W}(\btheta)$ that is close to $\bC_{W}(\btheta)$ in the
matrix 2-norm sense.  

\begin{rem}
The error $\| \tilde{\bC}_{W}(\btheta) - \bC_{W}(\btheta)\|_{2}$ will
be monotonically decreasing with respect to the smoothness of the
covariance function, the size of the degree of the multi-level basis
$\tilde f \geq f$ (accuracy parameter $\tilde{p} \geq p$) and the size
of $\tau$.  For a sufficiently large $\tau$ and $\tilde f$ the error
$\| \tilde{\bC}_{W}(\btheta) - \bC_{W}(\btheta)\|_{2}$ will be small
and the matrix $\tilde{\bC}_{W}(\btheta)$ becomes positive definite.

Now, the number of nonzeros of $\tilde{\bC}_{W}(\btheta)$ will
increase as we increase $\tau$ and $\tilde p$. To be able to
determine the size for $\tau$ and $\tilde p \geq p$ it is
helpful to derive an expression for the error $\|
\tilde{\bC}_{W}(\btheta) - \bC_{W}(\btheta) \|_{2}$ vs the number of
non zeros of $\tilde{\bC}_{W}(\btheta)$.

Error estimates can be derived for $\| \tilde{\bC}_{W}(\btheta) -
\bC_{W}(\btheta) \|_{2}$ with respect to the smoothness of the
covariance function, $\tilde{p}$ and $\tau$, but this is beyond the
scope of the present paper. In practice for the polynomial based model
${\cal Q}^d_{\tilde{f}}$ we set the level dependent criterion
parameter $\tau := 1$ and increase $\tilde{f}$ (and $\tilde{p}$) until
at least $\tilde{\bC}_{W}(\btheta)$ is positive definite. Moreover,
the sparse Cholesky factorization code in the Suite Sparse package
(\cite{Chen2008,Davis2009,Davis2005,Davis2001,Davis1999}) that is used
in this paper informs the user if the matrix is not positive definite.
\end{rem}

In \cite{Castrillon2013} the authors described how to apply a Kernel
Independent Fast Multipole Method (KIFMM) by \cite{ying2004} to compute
all the diagonal blocks $\tilde{\bC}^{i,i}_W(\btheta)$ for $i =
0,\dots,t$ in ${\cal O}(nt)$ computational steps to a fixed accuracy
$\varepsilon_{FM} > 0$. This approach can be easily extended to
compute all the blocks $\tilde{\bC}^{i,j}_W(\btheta)$ for $i,j =
-1,\dots,t$ in ${\cal O}(n(t + 1)^2)$.

\begin{rem}
The KIFMM by \cite{ying2004} is very flexible as it allows a large
class of covariance functions to be used including the exponential,
Gaussian and Mat\'{e}rn.  However, the computational efficiency is
mostly dependent on the implementation of the covariance function
(since the KIFMM computational cost is ${\cal O}(n)$) and the accuracy
parameter of the solver. For all the numerical experiments in this
paper the accuracy parameter is set to medium ($10^{-6}$ to $10^{-8}$)
or high ($10^{-8}$ or higher).

Due to the lack of a fast math C++ library for the Mat\'{e}rn
covariance function, we create a Hermite cubic spline interpolant of
the covariance function with the multithreaded Intel Math Kernel
Library (MKL) data fitting package. To generate a compact
representation of the interpolant we implement an $h$-adaptive mesh
generator in 1D such that the absolute error over the range $(0, 2.5]$
  is less than TOL. From \cite{Elden2004} given that the covariance
  function $\phi(r;\btheta) \in C^{4}(\R)$, $r \in \R$, on each mesh
  element (starting at $x_0 \in \R$) with length $h$ we can guarantee
  that the absolute error for the cubic Hermite interpolant is less
  than TOL if $\frac{h^4}{384} \max_{x \in
    [x_{0},x_{0}+h]}{\phi^{(4)}(x;\btheta)} < TOL$, where $\phi^{(4)}$
  refers to the fourth derivative with respect to $x$.  In this work
  we set $TOL = 5 \times 10^{-9}$.  Numerical test confirmed $TOL$
  accuracy for the Mat\'{e}rn covariance function with less than 200
  adaptive mesh nodes. This is sufficient for the numerical examples
  in this paper.
\label{multilevelcov:rem1}
\end{rem}

\begin{figure}[b!]
\begin{center}
\begin{tabular}{cc}
   \includegraphics[ width=3.0in,
     height=3in ]{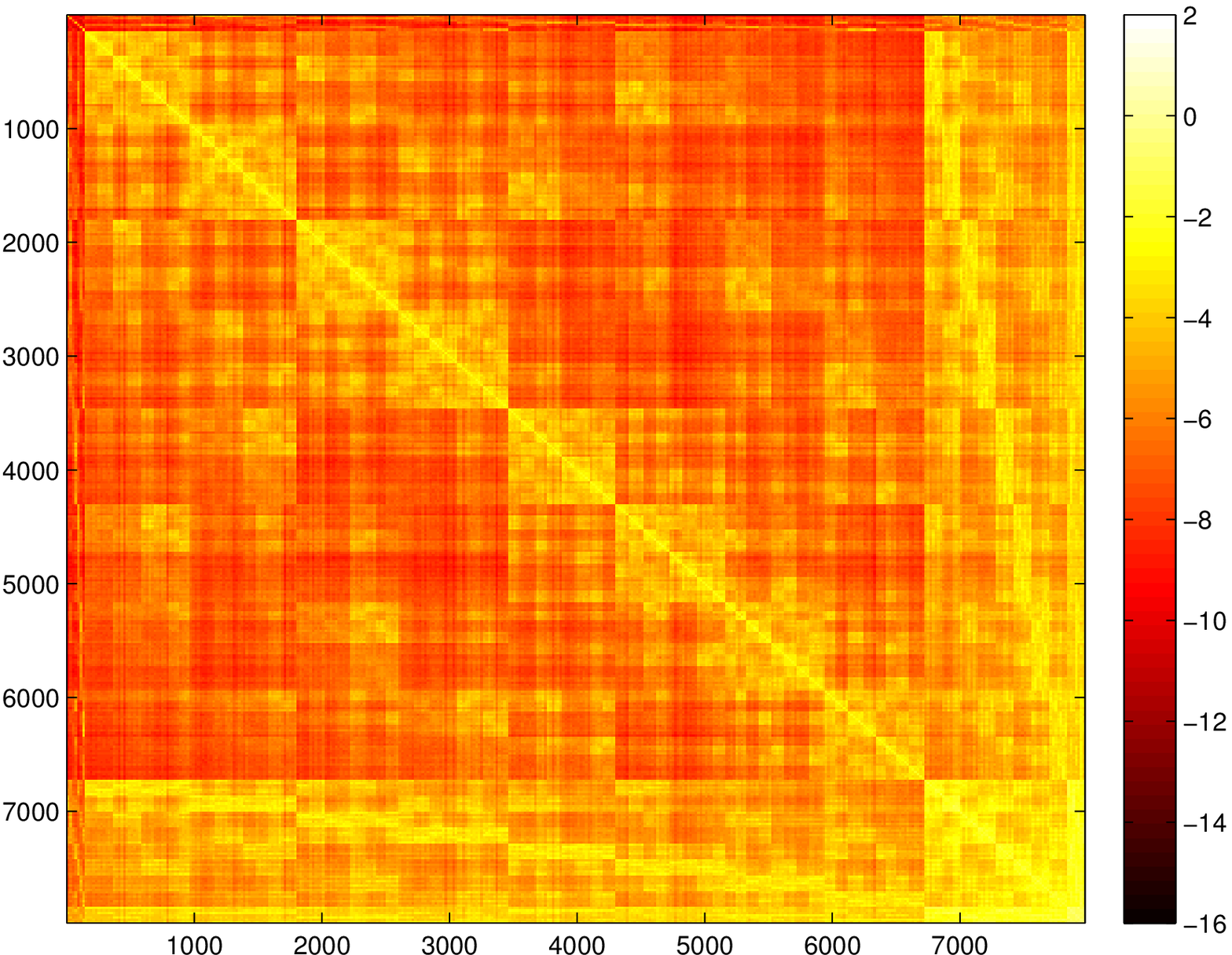} &
   \includegraphics[ width=3.0in, height=3in
   ]{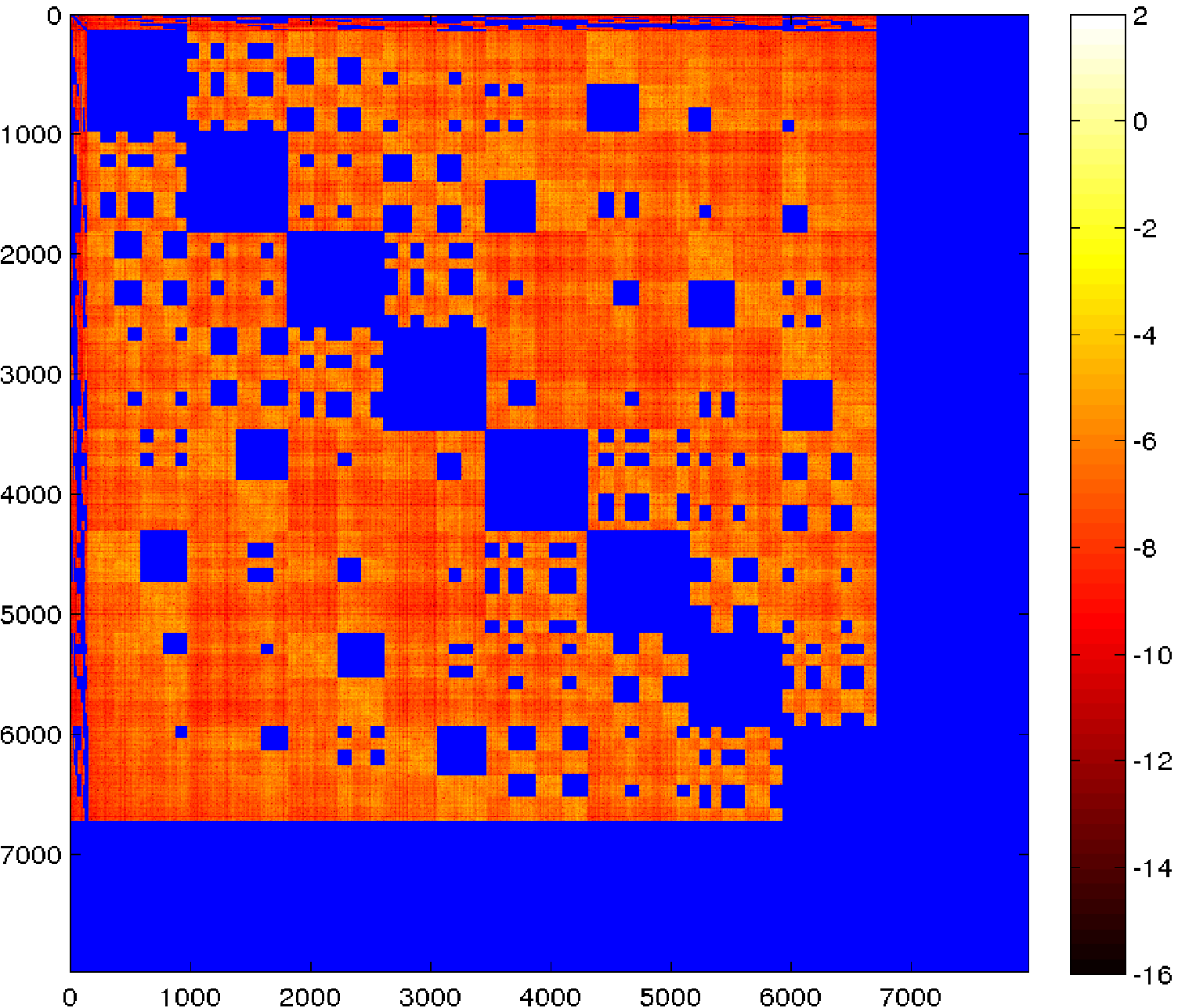}\\ (a) $\log_{10}{ abs( \bC_W(\btheta) )}$ & (b)
   Overlayed sparsity pattern in blue  \\
\end{tabular} 
\end{center}
  \caption{Sparsity pattern overlayed on $\bC_W(\btheta)$. Notice that
    most of the entries that are not covered by the blue boxes are around
    $10^{-7}$ times smaller in magnitude than the covered
    entries.}
\label{multilevelcov:fig4}
\end{figure}

In Figure \ref{multilevelcov:fig4} we show an example of a sparse
matrix produced for $\tau=1$ for $n = 8,000$ observation
locations sampled from a uniform distribution on the unit cube.
Notice that the entries of the matrix that are not covered by the
sparsity pattern are around $10^{-7}$ times smaller, implying the
hierarchical sparsity technique will lead to good accuracy.

The total sparsity for this example is 46\% (23\% since the matrix
$\tilde{\bC}_W(\btheta)$ is symmetric), however, the sparsity density
improves significantly as $n$ increases as we expect the number of
non-zero entries of $\tilde{\bC}_W(\btheta)$ to increase at most as
${\cal O}((t+2)^2n)$ with the number of observations $n$ (See 
\cite{Castrillon2013}). 

\begin{figure}[b!]
\begin{center}
   \includegraphics[ width=2.7in,
     height=3.3in ]{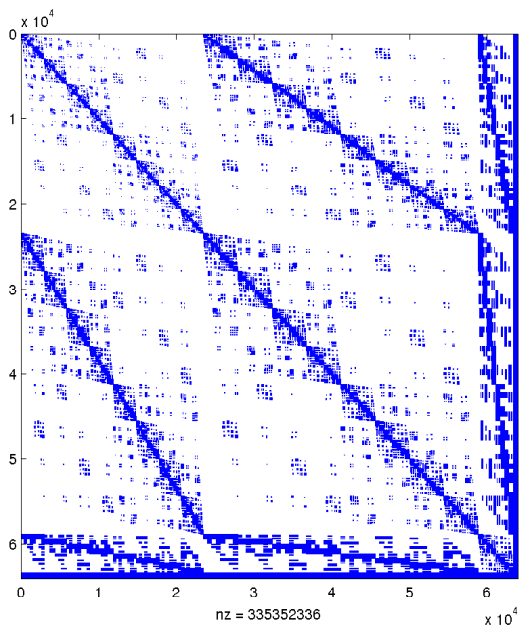}
\end{center}
  \caption{Sparsity pattern (8.2 \% non zeros) for $\tilde
    \bC_W$ with $\tau = 1$ and $n = 64,000$.}
\label{multilevelcov:fig6}
\end{figure}

In Figure \ref{multilevelcov:fig6} the sparsity pattern of the matrix
$\tilde{\bC}_W(\btheta)$ is shown for $n = 64,000$ observation
locations sampled from a uniform distribution on the unit cube.  For
this case the design matrix $\bM_f$ is constructed from $p = 20$
monomials (i.e. up to cubic polynomials) and $\tau = 1$.  The sparsity
of this example is $8.2\%$ (4.1 \% since the matrix
$\tilde{\bC}_W(\btheta)$ is symmetric).

\section{Multi-Level Estimator}
\label{multilevelestimator}

As the result section shows it is not necessary to compute the entire
sparse matrix $\bC_W(\btheta)$ to obtain a good estimate of the
covariance parameter $\btheta$. Due to the multi-resolution properties
of the MB we can construct a partial multi-resolution likelihood
function that is effective.

We can produce a series of multi-resolution likelihood functions
$\tilde{\ell}^i_{W}(\btheta)$, $i = -1,\dots t$ by applying the
partial transform $[\bW_t \T, \dots, \bW_i \T ]$ to the data $\bZ$,
thus
\begin{equation}
\tilde{\ell}^i_{W}(\btheta)
=-\frac{\tilde{n}}{2}\log(2\pi)-\frac{1}{2}\log
\det\{\tilde{\bC}_{W}^i(\btheta)\}
-\frac{1}{2}(\tilde{\bZ}^i_{W})\T\tilde{\bC}^i_{W}(\btheta)^{-1}\tilde{\bZ}^i_{W},
\label{Introduction:multilevelloglikelihoodreduced2}
\end{equation}
where $\tilde{\bZ}^i_W :=[\bW_t \T, \dots, \bW_i \T ] \T \bZ$,
$\tilde{n}$ is the length of $\tilde{\bZ}^i_W$ and
$\tilde{\bC}^i_{W}(\btheta)$ is the $\tilde{n} \times \tilde{n}$
upper-left submatrix of $\tilde{\bC}_{W}(\btheta)$.


\subsection{Computation of $\log{\det\{ \tilde{\bC}^i_{W}(\btheta)\}}$}

Since $\bC(\btheta)$ is symmetric positive definite from
\eqref{Introduction:eqn1} it can be shown that $\bC_{W}(\btheta)$ is
also symmetric positive definite.  It can also be shown that for a
sufficiently large $\tau$ and/or $\tilde{p}$ the matrix
$\tilde{\bC}^i_{W}(\btheta)$ will also be symmetric positive definite.
An approach to computing the determinant of
$\tilde{\bC}^i_{W}(\btheta)$ is to apply a sparse Cholesky
factorization technique such that $\bG\bG\T =
\tilde{\bC}^i_{W}(\btheta)$ where $\bG$ is a lower triangular
matrix. Since the eigenvalues of $\bG$ are located on the diagonal we
have that $\log \det \{\tilde{\bC}^i_{W}(\btheta)\} = 2 \sum_{i =
  1}^{\tilde{n}} \log{\bG_{ii}}$.

To reduce the fill-in of the factorization matrix $\bG$ we
apply the matrix reordering technique in Suite Sparse 4.2.1 package
(\cite{Chen2008,Davis2009,Davis2005,Davis2001,Davis1999}) 
with the Nested Dissection (NESDIS) function
package. The sparse Cholesky factorization is performed with the {\it lchol}
command from Suite Sparse 4.2.1 package.

Although in practice the combined NESDIS and sparse Cholesky
factorization is highly efficient, as shown by our numerical results,
a worse case complexity bound can be obtained.  For example, it can be
shown that by using the planar graph separation theorem (see
\cite{George1973}, \cite{Gilbert1987}) a worse case complexity of ${\cal
  O}(n^{3/2})$ and ${\cal O}(n (\log{n})^2 )$ storage is achieved in
2D. Similarly, the worse case complexity in 3D is ${\cal O}(n^{2})$.

\begin{exam}
Continuing Example \ref{multilevelcov:example1} we compute $\log{\det
  \{ \bC_{W}(\btheta)\}}$ and the approximation \linebreak
$\log{\det\{{\tilde{\bC}}_{W}(\btheta)\}}$ for $\tau = 0,1,2,\infty$
by applying the sparse Cholesky factorization. In
Table~\ref{multilevelcov:table1} we tabulated the absolute
$\varepsilon_{abs}:=|\log{\det\{{\tilde{\bC}}_{W}(\btheta)\}} -
\log{\det\{{\bC}_{W}(\btheta)\}}| $ and relative $\varepsilon_{rel}
:=\frac{\varepsilon_{abs}} { |\log \det \{\bC_{W}(\btheta)\}|} $
errors.  For $\tau = 0$ we obtain a very sparse matrix (4\% density),
but leads to a non-positive definite matrix, which is not valid for
the computation of the determinant. For $\tau = 1$ the matrix
$\tilde{\bC}_W(\btheta)$ becomes positive definite. As we increase
$\tau$ the approximation $\log{\det\{{\tilde{\bC}}_{W}(\btheta)\}}$
becomes more accurate. However, the density of
$\tilde{\bC}_{W}(\btheta)$ also increases.

\begin{table}[htp]
\caption{Log determinant errors comparisons between
   $\tilde{\bC}_{W}(\btheta)$ and ${\bC}_{W}(\btheta)$.}
\begin{center}
\begin{tabular}{c c c c c}
$\tau$ & density (\%) & $\log{\det\{{\tilde{\bC}}_{W}(\btheta)\}}$ &
$\varepsilon_{abs}$ 
& 
$\varepsilon_{rel}
$
\\ 
\hline
0 & $4$   & not positive definite & -- & -- \\
1 & $23$ & $-5.184354 \times 10^3$ & $2.23 \times 10^{-2}$ & $4.31
\times 10^{-6}$ \\
2 & $38$ & $-5.184332 \times 10^3$ & $3.78 \times 10^{-4}$ & 
$7.29 \times 10^{-8}$\\
$\infty$ &  $50$  & $-5.184331 \times 10^3$ & 0 & 0
\end{tabular}
\end{center}
\label{multilevelcov:table1}
\end{table}
\end{exam}

\subsection{Computation of $(\tilde{\bZ}^i_W)\T \tilde{\bC}^i_{W}(\btheta)^{-1}\tilde{\bZ}^i_{W}$}
\label{mutlilevelcomputation}

We have two choices for the computation of $(\tilde{\bZ}^i_W) \T \tilde
\bC^i_{W}(\btheta)^{-1} \tilde \bZ^i_{W}$. We can either use a Cholesky
factorization of $\tilde{\bC}^i_{W}(\btheta)$ or Preconditioned
Conjugate Gradient (PCG) coupled with a KIFMM.  The PCG choice
requires significantly less memory and allows more control of the
error. \corb{However, we already computed the sparse Cholesky factorization
of $\tilde{\bC}^i_{W}(\btheta)$ for the computation of the
determinant}. Thus we can use the same factors to compute $
(\tilde \bZ_{W}^i) \T\tilde{\bC}^i_{W}(\btheta)^{-1} \tilde \bZ^i_{W}$.


\section{Multi-Level Kriging}
\label{multilevelkriging}

An alternative formulation for obtaining the estimate $\hat \bZ(\bs_0)$ is
by solving the following problem
\begin{eqnarray}
\left( {{\begin{array}{*{20}c}
 \bC(\btheta) \hfill & \bM_f \hfill \\
 \bM_f\T \hfill & \0 \hfill \\
\end{array} }} \right)\left( {{\begin{array}{*{20}c}
 \hat \bgamma \hfill \\
 \hat \bbeta \hfill \\
\end{array} }} \right)=\left( {{\begin{array}{*{20}c}
 \bZ \hfill \\
 \0 \hfill \\
\end{array} }} \right).
\label{kriging:problem}
\end{eqnarray}
\corb{It is not hard to show that the solution of this problem
leads to equation \eqref{GLSbeta} and $\hat \bgamma(\btheta) =
\bC^{-1}(\btheta)\{\bZ - \bM_f \hat \hat \bbeta(\btheta)\}$ or
alternatively $\hat{\bbeta} = (\bM\T_f\bM_f)^{-1}\bM_f(\bZ -
\bC(\btheta)\hat{\bgamma})$.} The best unbiased predictor is
evaluated as
\begin{equation}
  \hat Z(\bs_0)
  =\bm(\bs_0)\T\hat \bbeta(\btheta)+\bc(\btheta)\T
  \hat \bgamma(\btheta)
\label{Kriging}
\end{equation}
and the Mean Squared Error (MSE) at the target point $\bs_0$ is given by
\begin{equation}
1 + 
\tilde{\bu}
\T
(\bM_f \T 
\bC(\btheta)^{-1} \bM_f )^{-1}
\tilde{\bu}
-\bc(\btheta)\T\bC^{-1}(\btheta)\bc(\btheta)
\label{Kriging:MSE}
\end{equation}
where $\tilde{\bu}\T := (\bM_f \bC^{-1}(\btheta)
\bc(\btheta) - \bm(\bs_0))$.

The computational cost for computing $\hat \bbeta(\btheta)$, $\hat
\bgamma(\btheta)$ and the MSE accurately using a direct method is
${\cal O}(n^3)$, which is unfeasible for large size problems. We
propose a much faster approach.

From \eqref{kriging:problem} we observe that $\bM_f\T \hat
\bgamma(\btheta) = \0$. This implies that $\hat{\bgamma} \in \R^{n}
\backslash {\cal P}^{p}(\mathbb{S})$ and can be uniquely rewritten as
$\hat{\bgamma} = \bW\T \bgamma_W$ for some $\bgamma_W \in
\R^{n-p}$. We can rewrite $\bC(\btheta) \hat \bgamma + \bM_f \hat
\bbeta = \bZ$ as  
\begin{equation}
\bC(\btheta) \bW\T \hat \bgamma_{W} + \bM_f \hat \bbeta =
       \bZ.
\label{Kriging:eqn1}
\end{equation}
Now apply the matrix $\bW$ to equation \eqref{Kriging:eqn1} and we
obtain $ \bW \{\bC(\btheta) \bW\T \hat \bgamma_{W} + \bM_f \hat
\bbeta\} = \bW \bZ.$ Since $\bW \bM_f = \0$ then $\bC_{W}(\btheta)
\hat \bgamma_{W} = \bZ_{W}$. Applying the preconditioner $\bD^{-1}_W
(\btheta)$, where $\bD_W (\btheta) := \diag(\bC_W (\btheta))$,
we have the system of equations $\bar{\bC}_{W}(\btheta)
\bar{\bgamma}_W (\btheta) = \bar{\bZ}_{W}$ where $\bar{\bgamma}_W
(\btheta) := \bD_W(\btheta) \bgamma_W (\btheta)$,
$\bar{\bC}_{W}(\btheta) := \bD^{-1}_W(\btheta) \bC_{W}(\btheta)
\bD^{-1}_W(\btheta)$ and $\bar{\bZ}_{W} := \bD^{-1}_W(\btheta)
\bZ_{W}$.

This system of equations is solved by a combination of a KIFMM and
PCG.  If $\bC_W(\btheta)$ and $\bD_W(\btheta)$ are symmetric positive
definite then an effective method to solve $\bar{\bC}_{W}(\btheta)
\bar{\bgamma}_W (\btheta) = \bar{\bZ}_{W}$ is the PCG method
implemented in PETSc by 
\cite{petsc-web-page,petsc-user-ref,petsc-efficient}. 

\begin{lem}
If the covariance function $\phi$ is positive definite, then the
matrix $\bD_W(\btheta)$ is always symmetric positive definite.
\label{multilevelkriging:lemma2}
\end{lem}

The matrix-vector products $\bC_W(\btheta) \bv$, where $\bv \in
\R^{n-p}$, are computed in ${\cal O}(n)$ computational steps to a
fixed accuracy $\varepsilon_{FM} > 0$.  The total computational cost
is ${\cal O}(kn(t+2))$, where $k$ is the number of iterations needed
to solve $\bar{\bC}_{W}(\btheta) \bar{\bgamma}_W (\btheta) =
\bar{\bZ}_{W}$ to a predetermined accuracy $\varepsilon_{PCG} > 0$.

It is important to point out that the introduction of a
preconditioner can degrade the performance of the PCG, in particular,
if the preconditioner is ill-conditioned. The accuracy of the PCG
method $\varepsilon_{PCG}$ has to be set such that the accuracy of the
{\it unpreconditioned} system $\bC_{W}(\btheta) \balpha_W (\btheta) =
\bZ_{W}$ is below a user given tolerance $\varepsilon > 0$.

\corb{We compute $\hat \bgamma = \bW\T \bgamma_W$ and $\hat{\bbeta} =
  (\bM\T_f\bM_f)^{-1}\bM_f \T (\bZ - \bC(\btheta)\hat{\bgamma})$ in at
  most ${\cal O}(np^2 + p^{3})$ computational steps. The matrix vector product
  $\bc(\btheta)\T \hat{\bgamma}(\btheta)$ is computed in ${\cal O}(n)$
  steps. Finally, the total cost for computing the estimate
  $\hat{\bZ}(\bs_0)$ from \eqref{Kriging} is ${\cal O}(np^2 + p^{3} +  kn(t+2))$}.

In Appendix C we show a procedure to compute the MSE fast.

\section{Numerical Study and Statistical Examples}
\label{numericalstudy}

In this section we test the numerical efficiency and accuracy of our
solver for computing the terms $\log{\det
  \{\tilde{\bC}^{i}_{W}(\btheta)\}}$ and $\bar{\bgamma}_W
(\hat{\btheta}) = \bar{\bC}^{-1}_{W}(\hat{\btheta}) \bar{\bZ}_{W}$ for
Mat\'ern covariances. Our results show that we are able to solve
problems of up to 128,000 observations and kriging up to 512,000 size
problems with good accuracy. Our approach is not limited to 128,000
for parameter estimation. This was the maximum we could test due to
the memory limitation on our workstation in creating observations
larger than 128,000. We now describe the data sets.

\begin{figure}[htpb]
\begin{center}
\begin{tabular}{c c c}
\includegraphics[width=2in, height=2in]{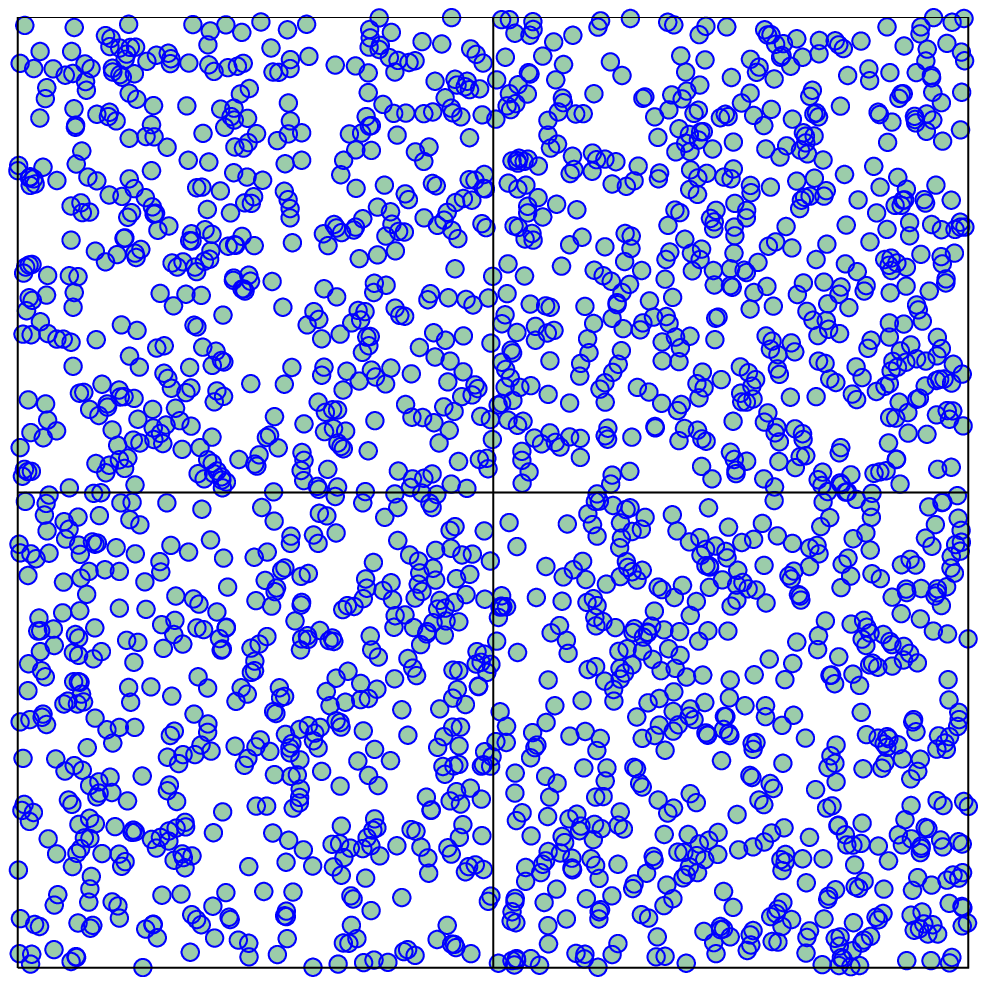} &
\includegraphics[width=2in, height=2in]{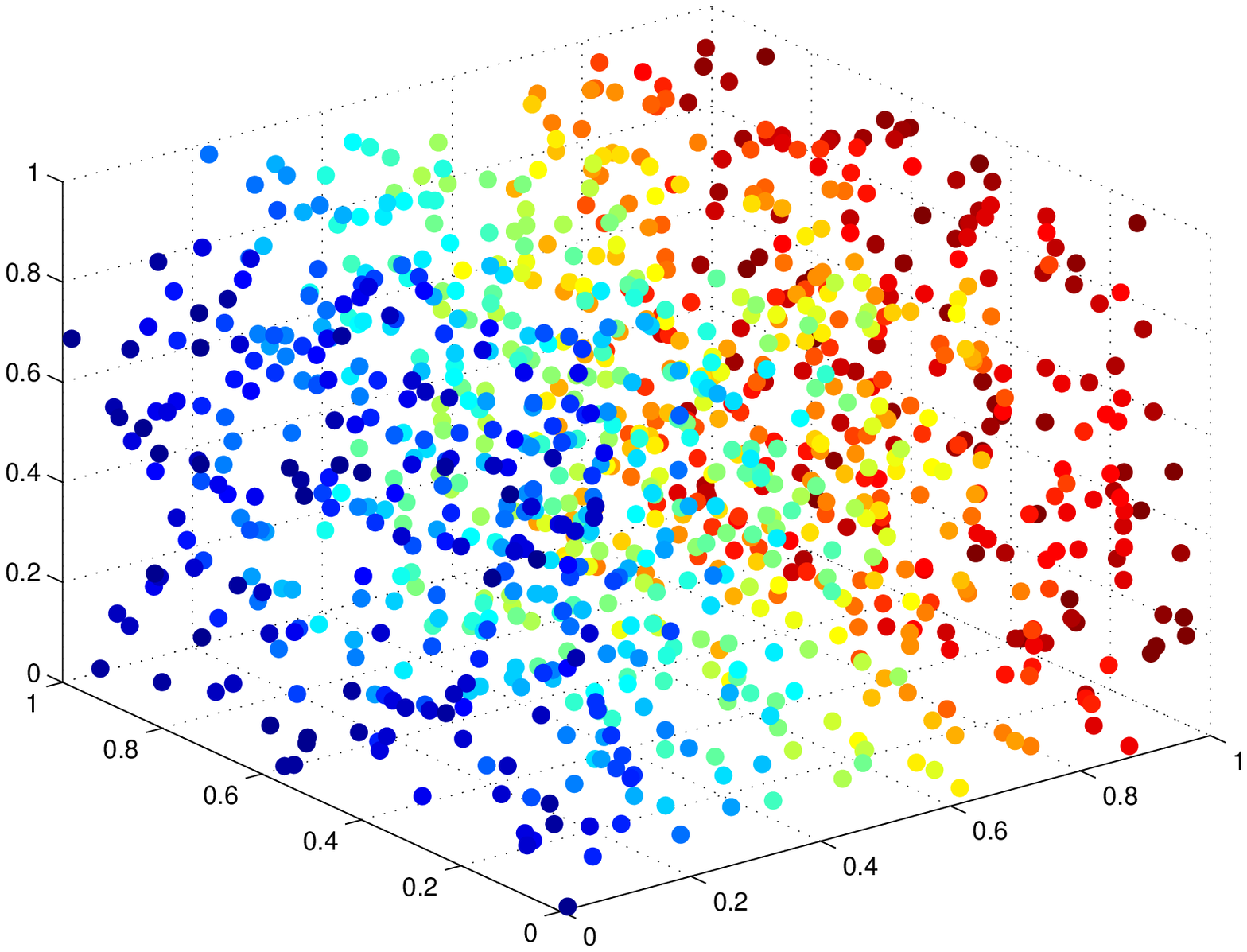} &
\includegraphics[width=2in, height=2in]{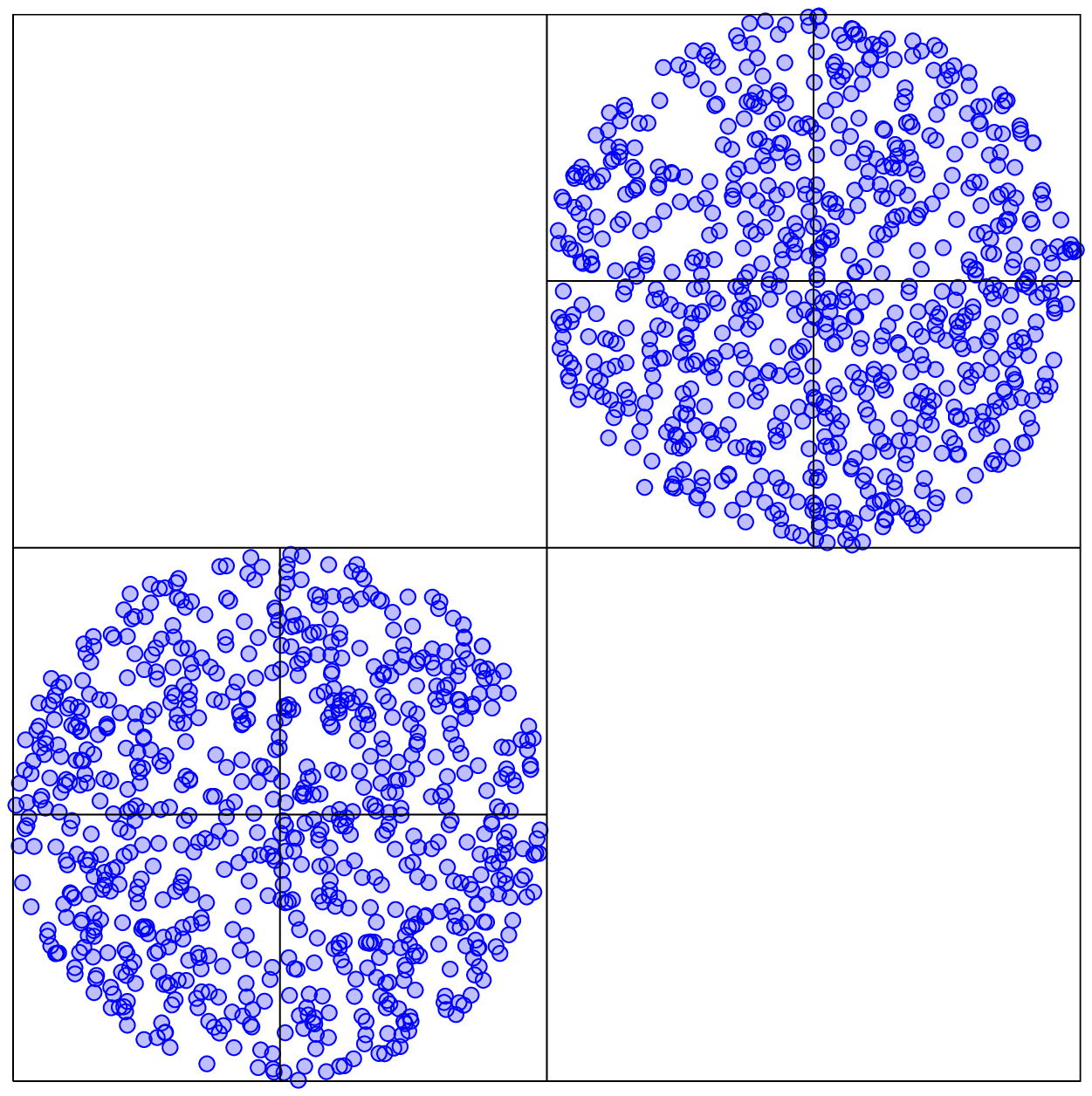} 
\\
(a) & (b) & (c)
\end{tabular}
\end{center}
\caption{Data set examples: (a) Data set \#1: One thousand observation
  points randomly generated from a uniform distribution on $[0,1]^2$.
  (b) Data set \#2: One thousand observation points randomly generated
  from a uniform distribution on $[0,1]^3$.  The color of the
  observation locations represent the distance along the right axis
  coordinate. (c) Data set \#3: Two disks of 1562 randomly generated
  observation locations. The disks are contained in a square
  but are represented in a multi-level representation.}
\label{SRBF}
\end{figure}

{\bf Data set \#1 and \#2:} The sets of observation locations
$\bS^d_{1}, \dots, \bS^d_{10}$ vary from 1,000 to 512,000 and we
assume that $\bS^d_{l} \subset \bS^d_{l+1}$ for $l = 1, \dots, 9$ for
$d = 2$ and $d = 3$.  The observations locations are sampled from a
uniform distribution over the unit square $[0,1]^2$ for $d = 2$ (data
set \#1) and for $[0,1]^3$ for $d = 3$ (data set \#2), as shown in
Figure \ref{SRBF}. The target points $\bs_0$ are set to 1000 random
points across the domain $[0,1]^2$ (data set \#1) and $[0,1]^3$(data
set \#2). We shall refer to $\bZ^d_1, \dots \bZ^d_{10}$ as the observation
values associated with $\bS^d_1, \dots \bS^d_{10}$.

{\bf Data set \#3:} We take the data set generated by $\bS^d_{9}$ for
$d = 2$ (256,000 observation points) and carve out two disks located
at (1/4,1/4) and (3/4,3/4) with radii 1/4. This generates 100,637
observation points; see Figure \ref{SRBF}(c) for an example with 1,562
observation points randomly extracted from the data set.

We now test our approach on the Mat\'{e}rn covariance function
$\phi(r;\btheta):=\frac{1}{\Gamma(\nu)2^{\nu-1}} \left(
\sqrt{2\nu}\frac{r}{\rho} \right)^{\nu} $ $K_{\nu} \left(
\sqrt{2\nu}\frac{r}{\rho} \right)$, where $\Gamma$ is the gamma
function and $K_{\nu}$ is the modified Bessel function of the second
kind. All results are executed on a single CPU (4 core Intel i7-3770
CPU @ 3.40GHz.) with Linux Ubuntu 13.04.

\vspace{-.4cm}

\subsection{Parameter Estimation}

In this section we present the results for data set \#1 and \#3
for the Mat\'{e}rn parameter estimation.

Suppose we have two realizations of the Gaussian spatial random field
$Z(\bs)$ with the Mat\'{e}rn covariance for data set \#1 (2D).  We set
$f = 3$ ($p = 10$) and $\tilde{f} = 4,5,6$ (corresponding to
$\tilde{p} = 15,21,28$) and fix the covariance parameters to $\btheta=(\nu,\rho)
= (3/4,1/6)$.  Two realizations $\bZ^2_6$ ($n = 64,000$) and $\bZ^2_7$
($n = 128,000$) are generated from these parameters. For each
observation values $\bZ^2_{6}$ and $\bZ^2_{7}$ (and locations) apply
the transformation $\bW$ to compute $\bZ^2_{W,6}$ and $\bZ^2_{W,7}$
and solve the optimization problems $\hat{\btheta}_{i} : =
  \argmax_{\btheta \in (0,\infty) \times (0,\infty)}
  \hat{\ell}_{W}(\bZ^2_{W,i},\btheta)$ for $i = 6$ and $7$.

\begin{table}[b!]
\caption{Estimation results for data sets \#1 and \#3.  The
  observation data is generated with covariance parameters $\nu=0.75$
  and $\rho=1/6$.  The degree of the model is $f = 3$ (cubic), which
  gives $p = 10$ monomials and we set $\tau :=1$.  \corb{ For all the
    experimental runs the number of MB levels is $t = 6$.}  The first
  column is the size of the problem. Columns 2 and 3 are the
  parameters that are chosen to build the MB and the multi-level
  estimator.  \corb{Column 4 is the corresponding $\tilde p$ for
    $\tilde f$ and the maximum level.  Columns 5 and 6 are the errors
    of $\hat \nu - \nu$ and $\hat \rho - \rho$.  Column 7 is the
    percentage of non-zeros of the Cholesky factors. Column 8 is
    self-explanatory.  Column 9 and 10 are the approximate wall clock
    computational time (in seconds) needed to compute
    $\tilde{\bC}^{i}_W(\btheta)$ and to perform Cholesky factorization
    respectively.} The total time for each Newton iteration is about
  $t_{cons} (s) + t_{chol} (s)$.  For each problem it takes about 50
  Newton iterations to converge.  \corb{Estimation results for data
    set \#3 with observation data generated with covariance parameters
    $\nu=1.25$ and $\rho=1/6$.}  }
\begin{center}
Estimation results for data set \#1 (2D). 
\\
\corb{
\begin{tabular}{ r r r | r || r r r r r r r}
\multicolumn{1}{c}{$n$} &
\multicolumn{1}{c}{$\tilde f$} &
\multicolumn{1}{c}{$i$} &
\multicolumn{1}{c}{$\tilde p$} & 
\multicolumn{1}{c}{$\hat{\nu} - \nu$} &
\multicolumn{1}{c}{$\hat{\rho} - \rho$} &
\multicolumn{1}{c}{$nz(\bG) (\%)$} &
\multicolumn{1}{c}{size($\tilde{\bC}^i_W$)} &
\multicolumn{1}{c}{$t_{cons} (s)$} &
\multicolumn{1}{c}{$t_{chol} (s)$} &
 \\ 
\hline \hline
64,000 &  6 &  6 & 28 &  -0.0759 &  0.0333 & 8.9 &   23  &  14 & 0  \\
64,000 &  6 &  5 & 28 &   0.0182 & -0.0132 & 1.7 & 35328 &  40 & 1  \\
64,000 &  6 &  4 & 28 &  -0.0043 &  0.0046 & 4.5 & 56832 & 230 &11  \\
64,000 &  6 &  3 & 28 &  -0.0049 &  0.0048 & 10.7 & 62208 & 961 &65  \\
\hline
64,000 &  5 &  6 & 21 &  0.0071 &  -0.0146 & 0.4 &   810 &  13 & 0  \\
64,000 &  5 &  5 & 21 &  0.0037 &  -0.0027 & 1.7 & 42496 &  43 & 2  \\
64,000 &  5 &  4 & 21 & -0.0030 &   0.0048 & 3.7 & 58624 & 220 &12  \\
64,000 &  5 &  3 & 21 & -0.0048 &   0.0046 & 7.0 & 62656 & 750 &32  \\
\hline
64,000 &  4 &  6 & 15 & -0.0080 & 0.0098 & 0.2 &  7749 &  13 & 0  \\
64,000 &  4 &  5 & 15 & -0.0047 & 0.0043 & 2.0 & 48640 &  53 & 3  \\
64,000 &  4 &  4 & 15 & -0.0068 & 0.0062 & 3.4 & 60160 & 161 & 9  \\
64,000 &  4 &  3 & 15 & -0.0051 & 0.0048 & 4.4 & 63040 & 550 &15  \\
\hline \hline 
128,000 &  6 &  6 & 28 &  0.0010 & -0.0011 & 0.3 & 17179 & 75 & 0 \\
128,000 &  6 &  5 & 28 &  0.0025 & -0.0020 & 2.1 & 99328 & 350 &13 \\
128,000 &  6 &  4 & 28 & -0.0002 &  0.0005 & 4.0 &  120832 & 1200 & 70 \\
\hline
128,000 &  5 &  6 & 21 & -0.0010 &  0.0015 & 0.5 &   42154 &   80 &  0 \\
128,000 &  5 &  5 & 21 &  0.0004 & -0.0002 & 1.9 &  106496 &  300 & 14 \\
128,000 &  5 &  4 & 21 & -0.0016 &  0.0017 & 3.3 &  122624 & 1000 & 50 \\
\end{tabular} 
}
\\
\vspace{5mm}
Estimation results for data set \#3 (2D). \\
\corb{
\begin{tabular}{ r r r | r || r r r r r r r}
\multicolumn{1}{c}{$n$} & 
\multicolumn{1}{c}{$\tilde f$} &
\multicolumn{1}{c}{$i$} &
\multicolumn{1}{c}{$\tilde{p}$} &
\multicolumn{1}{c}{$\hat{\nu} - \nu$} &
\multicolumn{1}{c}{$\hat{\rho} - \rho $} &
\multicolumn{1}{c}{$nz(\bG) (\%)$} &
\multicolumn{1}{c}{size($\tilde{\bC}^i_W$)} &
\multicolumn{1}{c}{$t_{cons} (s)$} &
\multicolumn{1}{c}{$t_{chol} (s)$} &
 \\ 
\hline
100,637 &  6 &  6  & 66 &  0.0548 & -0.0237 & 0.5 &  2613 &  60 &  0 \\
100,637 &  6 &  5  & 66 & -0.0031 & 0.0020 & 3.1 & 72231 & 600 & 12 \\
\end{tabular}
}
\vspace{5mm}
\\

\end{center}
\label{SRBF:table6}
\end{table}
The optimization problem from the log-likelihood function
\eqref{Introduction:multilevelloglikelihood} is solved using {\it
  fminsearch} from the optimization toolbox in MATLAB with the local
minimizer search for $\nu$ in the interval $[1/2,5/4]$ and $\rho$ in
the interval $[1/7,1/5]$.  We set the parameter criterion to $\tau :=
1$, and the {\it fminsearch} tolerance is set to $10^{-3}$.  In Table
\ref{SRBF:table6} we tabulate the results for the parameter estimates
$\hat \nu$ and $\hat \rho$ for different problem sizes of data sets
\#1 and \#3 for the user defined parameters $\tilde f$ for the
construction of the MB and $i$ for the construction of the matrix
$\tilde \bC^{i}_W$. We notice that the estimates of $(\nu,\rho)$ seem to
approach the actual values as we increase the number of
observations. In particular, for $n = 128,000$ the estimate
$(\hat{\nu},\hat{\rho}) := (0.7498,0.1672)$ is very close to the
actual noise model parameters $(\nu, \rho) = (0.75,1/6)$ of the
covariance function. We observe that as we increase the number of levels
(i.e. decrease $i$) in the covariance matrix the absolute error decays
until it stagnates, usually by the time that the covariance matrix is
for two levels.  We also report the wall clock times (i.e. actual time
it took the executable to run, not to be confused with CPU time that
is unreliable as a measure) for computing each Newton iteration.  The
total number of Newton iterations is approximately~50.

In Table \ref{SRBF:table6} the results for parameter estimation
with data set \#3 are tabulated.  The realization is obtained from the
Gaussian random field $Z(\bs)$ with $n = 100,637$, $\nu = 1.25$ and
$\rho = 1/6$. For this case the absolute error is 0.25\% for the
estimate $\hat{\nu}$ and 0.04\% for $\hat{\rho}$.

\begin{table}[b!]
\caption{ (a) Statistical results for data set \#1 with observation
  data generated with covariance parameters $\nu=0.75$ and $\rho=1/6$.
  \corb{The parameter $M = 100$ is the number of realizations of the
    stochastic model, $\bbE_M [\hat{\nu} - \nu]$ is the bias of $M$
    estimates of $\nu$ and $std_M [\hat{\nu}]$ is the standard
    deviation of $M$ estimates of $\nu$.}  As $i$ is reduced from $t$
  to $t-1$ there is a significant drop in $std_M
  [\hat{\nu}]$. However, for $i < t - 1$ the standard deviation of the
  estimates does not improve significantly. This indicates that a good
  estimate can be obtained for $i = t-1$ and there is not much gain in
  reducing $i$, which increases the computational cost in computing
  $\tilde{\bC}^{i}_W$. }

\begin{center}
Statistical results for data set \#1 with multiple realizations.
\\
\corb{
\begin{tabular}{ r r r | r r || r r r r r r}
\multicolumn{1}{c}{$n$} & 
\multicolumn{1}{c}{$\tilde f$} &
\multicolumn{1}{c}{$i$} &
\multicolumn{1}{c}{$\tilde{p}$} &
\multicolumn{1}{c}{$t$} &
\multicolumn{1}{c}{$\bbE_M [\hat{\nu} - \nu]$} &
\multicolumn{1}{c}{$\bbE_M [\hat{\rho} - \rho]$} &
\multicolumn{1}{c}{$std_M [\hat{\nu}]$} &
\multicolumn{1}{c}{$std_M [\hat{\rho}]$} &
 \\ 
\hline
\hline
32,000 &  4 &  5  & 15 &  5 & $-6.0 \times 10^{-4}$  
& 
$  1.0 \times 10^{-3}$  
& $1.3 \times 10^{-2}$ 
& $1.0 \times 10^{-2}$ \\
32,000 &  4 &  4  & 15 & 5 & $-7.2 \times 10^{-4}$  
& $7.0 \times 10^{-4}$ 
& $5.9  \times 10^{-3}$ & 
$4.5 \times 10^{-3}$
      \\
32,000 &  4 &  3  & 15 & 5 & $-7.0 \times 10^{-4}$  
& $6.0 \times 10^{-4}$ 
&
$5.6  \times 10^{-3}$ &
$4.0   \times 10^{-3}$ \\
\hline
64,000 &  4 &  6  & 15 & 6 & $6.8 \times 10^{-4}$ 
& $1.1 \times 10^{-3}$ 
& $2.0  \times 10^{-2}$ &
$1.9   \times 10^{-2}$ \\
64,000 &  4 &  5  & 15 & 6 & 
$7.4 \times 10^{-4}$ 
& $-5.6 \times 10^{-4}$ 
& $5.4  \times 10^{-3}$ &
 $4.6   \times 10^{-3}$ \\
64,000 &  4 &  4  & 15 & 6 & 
$2.5 \times 10^{-4}$ 
& $-1.7 \times 10^{-4}$ 
& $3.9  \times 10^{-3}$ &
 $3.3   \times 10^{-3}$ \\
\hline
128,000 &  6 &  6  & 28 & 6 &
$-1.3 \times 10^{-3}$ 
& $1.5 \times 10^{-3}$ 
& $8.3  \times 10^{-3}$ &
 $7.7   \times 10^{-3}$ \\
128,000 &  6 &  5  & 28 & 6 & 
$-6.2 \times 10^{-4}$ 
& $6.5 \times 10^{-4}$ 
& $3.7  \times 10^{-3}$ &
 $3.3   \times 10^{-3}$ \\
\end{tabular}
}
\end{center}
\label{SRBF:table7}
\end{table}

In Table \ref{SRBF:table7} we generate $M=100$ realizations of the
stochastic model for data set \#1, to analyze the mean and standard
deviation of the Mat\'{e}rn covariance parameter estimates. The mean
estimate $\bbE_M [\hat{\nu}]$ refers to the mean of $M$ estimates
$\hat{\nu}$ for the the $M$ realizations of the stochastic
model. Similarly, $std_M [\hat{\nu}]$ refers to the the standard
deviation of the $M$ realizations. We observe that the mean appears to
approach the covariance parameters as we decrease $i$. As $i$ is
reduced from $t$ to $t-1$ there is a significant drop in the term
$std_M [\hat{\nu}]$. However, for $i < t - 1$ the standard deviation
$std_M [\hat{\nu}]$ does not improve significantly. Therefore, there
is not much gain in improving the estimate by decreasing $i$, which
increases the computational cost in computing $\tilde{\bC}^{i}_W$.

\subsection{Numerical examples for computing $\bC_{W}(\btheta)^{-1}\bZ_{W}$
and Kriging}

We test our approach for solving the system of equations
$\bar{\bC}_{W}(\btheta)\bar{\bgamma}_{W} = \bar \bZ_{W}$ (that we have
to solve to obtain the kriging predictor) on the data sets \#1 and
\#2. We also include results showing the kriging prediction error
between the multi-level and direct methods.

We first test the PCG method with data set \#2 (3D) on three test
cases: (a) $\btheta_a = (\nu,\rho)= (3/4,1/6)$, (b) $\btheta_b =(1,1/6)$ and
(c) $\btheta_c = (5/4,1/6)$. The value $\rho = 1/6$ gives us an
approximate decay of $5 \%$ (which is reasonable in practice) from the
center of the cube along each dimensional axis.  The PCG relative
error tolerance $epsilon_{PCG} > 0$ is set to a value that leads
to a relative error $\epsilon = 10^{-3}$ of the {\it unpreconditioned}
system $\bC_{W}(\btheta) \hat \bgamma_{W} = \bZ_{W}$.

In Table \ref{SRBF:table1} we report the total wall clock times and
iterations for computing $\hat{\bbeta}$, $\hat{\bgamma}$ and the
target $\hat \bZ(\bs_0)$ for data set \#2 (3D) with the Mat\'{e}rn
covariance function . The polynomial accuracy of the model is set to
cubic ($f = 3$, $p = 20$) and the accuracy parameter $\tilde{p}$ is
set to 20 (which corresponds to $\tilde{f} = 3$).  We look at three
cases: For (a) ($\btheta_a = (3/4,1/6)$) we set the KIFMM accuracy
to medium and the number of iterations increase as ${\cal
  O}(n^{0.58})$. For (b) ($\btheta_b = (1,1/6)$) we set the KIFMM
accuracy to medium and the number of iterations increases as ${\cal
  O}(n^{0.58})$. For (c) ($\btheta_c = (5/4,1/6)$) we set the KIFMM
accuracy to high and the number of iterations increases as ${\cal
  O}(n^{0.77})$.

\corb{In Table \ref{SRBF:table1} we also report the number of
  iterations needed for solving $\bC^{-1}(\btheta)\bZ$ with $10^{-3}$
  accuracy with a CG method.} In this case the number of iterations is
about 10 times larger than the multi-level version.  Moreover, for
solving the kriging problem (e.g. equation \eqref{GLSbeta}), $p$ such
problems have to be solved. Thus, it is at least about 200 times
faster since $p = 20$ for this case. An alternative is to solve
\eqref{kriging:problem}. However, in general it is not positive
definite. The matrix is highly ill-conditioned also making it
difficult to solve with an iterative solver such as generalized
minimal residual method (see \cite{Castrillon2013}).

\begin{table}[t!]
\caption{Diagonal pre-conditioned results for computing
  $\bar{\bC}_{W}(\btheta) \bar{\bgamma}_{W} = \bar{\bZ}_{W}$ for data
  set \#2 (3D) with the Mat\'{e}rn covariance $\btheta =
  (\nu,\rho)$. We look at three cases: (a) $\btheta_a = (3/4,1/6)$ (b)
  $\btheta_b = (1,1/6)$ and (c) $\btheta_c = (5/4,1/6)$.  The relative
  error of the residual of the unpreconditioned system is set to
  $\varepsilon = 10^{-3}$. The KIFMM is set to medium accuracy for (a)
  and (b), and set to high accuracy for (c). The second column is the
  number of iterations needed to obtain $10^{-3}$ relative error of
  the unpreconditioned system with $\bC_W(\btheta)$. We denote as
  itr($\bC_W$) as the number of CG iterations needed for convergence
  until the desired accuracy is achieved.  The third column is the
  number of iterations for solving $\bC^{-1}(\btheta)\bZ$ with $10^{-3}$
  accuracy.  The fourth column is the residual tolerance needed for
  convergence of $10^{-3}$ relative error for the unpreconditioned
  system. The fifth column presents the wall clock times for
  initialization (basis construction and preconditioner
  computation). The PCG iteration wall clock times for $\bC_W$ are
  given in the sixth column. The last column presents the total wall
  clock time to compute $\bar{\bgamma}_{W} =
  \bar{\bC}_{W}(\btheta)^{-1} \bar \bZ_{W}$.}
\begin{center}
\corb{(a) $\btheta_a = (3/4,1/6)$, $d = 3$, $f = 3$ ($p = 20$), $\tilde{f} = 3$ ($\tilde{p} = 20$)} \\
\begin{tabular} { r r r r r r r }
  \multicolumn{1}{c}{$n$} & itr($\bC_W$) & itr($\bC$) & \multicolumn{1}{c}{$\varepsilon_{PCG}$}  & Diag. (s) & Itr (s) & Total (s) \\
  \hline
  16,000  &    166 & 1296 & $1.02 \times 10^{-4}$   &      80 &     113 &     193 \\
  32,000  &    247 & 3065 & $9.88 \times 10^{-5}$   &     215 &     321 &     536 \\
  64,000  &    372 & 5517 & $1.00 \times 10^{-4}$   &     665 &    1226 &    1891 \\
  128,000 &    547 &    - & $4.84 \times 10^{-5}$   &    2060 &    3237 &    5397 \\
  256,000 &    847 &    - & $5.00 \times 10^{-5}$   &    5775 &    9885 &   15660 \\
  512,000 &   1129 &    - & $3.74 \times 10^{-5}$   &   17896 &   33116 &   51012 \\
\end{tabular}\\
\corb{(b) $\btheta_b = (1,1/6)$, $d = 3$, $f = 3$ ($p = 20$), $\tilde{f} = 3$ ($\tilde{p} = 20$)} \\
\begin{tabular} { r r r r r r r }
  \multicolumn{1}{c}{$n$} & itr($\bC_W$) & itr($\bC$) & \multicolumn{1}{c}{$\varepsilon_{PCG}$}  & Diag. (s) & Itr (s) & Total (s) \\
  \hline
  16,000  &    293 &  2970 &$8.23 \times 10^{-5}$   &     79  &     198 &     277 \\
  32,000  &    470 &  7786 &$8.18 \times 10^{-5}$   &    213  &     607 &     820 \\
  64,000  &    760 & 15808 &$7.09 \times 10^{-5}$   &    662  &    2495 &    3157 \\
  128,000 &   1167 &     - &$3.00 \times 10^{-5}$   &   2050  &    7109 &    9159 \\
  256,000 &   1961 &     - &$3.27 \times 10^{-5}$   &   5789  &   22878 &   28667 \\  
\end{tabular}\\
\corb{(c) $\btheta_c = (5/4,1/6)$, $d = 3$,
$f = 3$ ($p = 20$), $\tilde{f} = 3$ ($\tilde{p} = 20$)} \\
\begin{tabular} { r r r r r r r }
  \multicolumn{1}{c}{$n$} & itr($\bC_W$) & itr($\bC$) & \multicolumn{1}{c}{$\varepsilon_{PCG}$}  & Diag. (s) & Itr (s) & Total (s) \\
  \hline
  16,000  &    500 &  5953 & $6.27 \times 10^{-5}$   &     138 &     580 &    718   \\
  32,000  &    827 & 17029 & $7.29 \times 10^{-5}$   &     346 &    1574 &   1920  \\
  64,000  &   1567 & 37018 & $4.45 \times 10^{-5}$   &     910 &    6474 &   7384  \\
  128,000 &   2381 &     - & $2.23 \times 10^{-5}$   &    3974 &   25052 &  29026  \\
  256,000 &   4299 &     - & $2.61 \times 10^{-5}$   &   10322 &   72374 &  82696   \\
\end{tabular}\\
\end{center}
\label{SRBF:table1}
\end{table}

In Table \ref{SRBF:table2} the results for computing $\hat{\bbeta}$,
$\hat{\bgamma}$ and $\hat Z(\bs_0)$ for 1000 target points $\bs_0$ for
data set \#1 (2D) with the Mat\'{e}rn covariance function are
tabulated. We have three test cases: (a) $\btheta_a =(\nu,\rho)=
(1/2,1/6)$ (note for this case we obtain an exponential covariance
function), (b) $\btheta_b = (3/4,1/6)$, and (c) $\btheta_c = (1,1/6)$.
For (a) and (b) the KIFMM accuracy is set to medium. For (c) the KIFMM
accuracy is set to high. For this case the relative residual accuracy
for the unpreconditioned system is fixed at $10^{-2}$.

\corb{We note that for this case the results are even more impressive
  than for the 3D case}. For Table \ref{SRBF:table2} (c) the CG solver
stagnated and we terminated the iteration after 100,000. At this point
the matrix $\bC(\btheta)$ is highly ill-conditioned. In contrast, with
$\bC_W(\btheta)$ we are still able to solve the problem even for
128,000 size problem.

\begin{table}[t!]
\caption{Diagonal pre-conditioned results for computing
  $\bar{\bC}_{W}(\btheta)\bar{\bgamma}_{W} = \bar{\bZ}_{W}$ for the
  Mat\'{e}rn covariance with $\btheta = (\nu,\rho)$ for data set \#1 (2D). We look at three
  cases: (a) $\btheta_a = (3/4,1/6)$, (b) $\btheta_b = (1,1/6)$ and
  (c) $\btheta_c = (5/4,1/6)$.  The relative error of the residual of
  the unpreconditioned system is set to $\varepsilon = 10^{-2}$. The
  KIFMM is set to medium accuracy for (a), (b) and set to high
  accuracy for (c).}
\begin{center}
\corb{
(a) $\btheta_{a} = (1/2,1/6)$, $d = 2$, $f = 3$  ($p = 10$), $\tilde{f} = 12$  ($\tilde{p} = 91$)
}\\
\begin{tabular} { r r r r r r r }
  \multicolumn{1}{c}{$n$} & itr($\bC_W$) & itr($\bC$) & \multicolumn{1}{c}{$\varepsilon_{PCG}$}  & Diag. (s) & Itr (s) & Total (s) \\
  \hline
  16,000  &    330 & 3603 & $2.39 \times 10^{-3}$    &     246 &     115 &   361   \\
  32,000  &    333 & 5429 & $1.39 \times 10^{-3}$    &     750 &     251 &  1001   \\
  64,000  &    455 & 8152 & $1.32 \times 10^{-3}$    &    1947 &     589 &  2536   \\
  128,000 &    564 &    - & $7.10 \times 10^{-4}$    &    5570 &    1577 &  7147   \\
  256,000 &    619 &    - & $9.78 \times 10^{-4}$    &   15266 &    3065 & 18331   \\
  512,000 &   1230 &    - & $4.50 \times 10^{-4}$    &   42254 &   13101 & 55355   \\
\end{tabular}\\
\corb{(b) $\btheta_b = (3/4,1/6)$, $d = 2$, $f = 14$ ($p = 120$), $\tilde{f} = 14$ ($\tilde{p} = 120$)} \\
\begin{tabular} { r r r r r r r }
  \multicolumn{1}{c}{$n$}  & itr($\bC_W$) & itr($\bC$) & \multicolumn{1}{c}{$\varepsilon_{PCG}$}  & Diag. (s) & Itr (s) & Total (s) \\
  \hline
  16,000  &   965  &  26795 & $2.78 \times 10^{-3}$    &     370 &     397 &    767  \\ 
  32,000  &  1110  &  41079 & $2.04 \times 10^{-3}$    &    1125 &    1061 &   2186  \\
  64,000  &  2239  &  82166 & $1.35 \times 10^{-3}$    &    2892 &    3714 &   6606  \\
  128,000 &  3443  &      - & $1.09 \times 10^{-3}$    &    8268 &   13130 &  21398  \\
  256,000 &  4557  &      - & $7.63 \times 10^{-4}$    &   23175 &   30302 &  53477  \\
\end{tabular}\\
\corb{
(c) $\btheta_c = (1,1/6)$, $d = 2$, $f = 14$ ($p = 120$), $\tilde{f}
= 14$ ($\tilde{p} = 120$) }\\
\begin{tabular} { r r r r r r r }
  \multicolumn{1}{c}{$n$} & itr($\bC_W$) & itr($\bC$) & \multicolumn{1}{c}{$\varepsilon_{PCG}$}  & Diag. (s) & Itr (s) & Total (s) \\
  \hline
  16,000  &  2710  &  $>100,000$ &  $1.90 \times 10^{-3}$  &     553 &    1844 &    2397  \\
  32,000  &  4261  &          -  &  $1.43 \times 10^{-3}$  &    1522 &    5713 &    7235  \\
  64,000  &  8801  &          -  &    $1.00 \times 10^{-4}$  &    5022 &   23785 &   28807  \\
  128,000 & 14405  &          -  &  $7.28 \times 10^{-4}$  &   12587 &   75937 &   88524 \\
\end{tabular}\\
\end{center}
\label{SRBF:table2}
\end{table}

The diagonal preconditioner we use is one of the simplest. We plan to
extend this approach to more sophisticated preconditioners such as
block Symmetric Successive OverRelaxation (SSOR) (see
\cite{Castrillon2013}) in the future.

The residual errors are then propagated to the final estimate
$Z(\bs_0)$ around the same magnitude. However, as a final experiment
in Table \ref{SRBF:table3} we tabulate the relative $l_2$ error
between the multi-level kriging approach and the direct method for
data set \#2 with exponential covariance function $\exp(-\theta r)$,
where $\theta = 5.9915$ and $f = 3$. The PCG tolerance is set to
$10^{-5}$ and $\tilde f = 3$. Notice that the error increases with
$n$. This is expected since the unpreconditioned system error will
degrade.
\begin{table}[h!]
\caption{Tabulation of the kriging estimate relative $l_2$ error
  between the multi-level kriging approach and the direct method for
  Data Set \#1 (3D) for 1000 target points. The covariance function is
  $\exp(-\theta r)$, where $\theta = 5.9915$. The polynomial bias term in
  the Gaussian model is set to $f = 3$ ($p = 20$). The solver
  tolerance is set to $\varepsilon_{PCG} = 10^{-5}$ and accuracy
  parameter is set to $\tilde p = 20$ ($\tilde f = 3$). }
 \begin{center}
 \begin{tabular}{ r c }
 \multicolumn{1}{c}{$n$} & 
 \multicolumn{1}{c}{$l_{2}$ Relative Error} \\
 \hline 
 1,000  &   $1.53 \times 10^{-6}$ \\
 2,000  &   $6.71 \times 10^{-5}$ \\
 4,000  &   $6.42 \times 10^{-5}$ \\
 8,000  &   $1.01 \times 10^{-4}$ \\
 16,000 &   $9.14 \times 10^{-5}$ \\
 32,000 &   $1.05 \times 10^{-4}$ \\
 \end{tabular}
 \end{center}
 \label{SRBF:table3}
\end{table}

\section{Conclusions}
\label{multilevelconclusions}

 In this paper we developed a multi-level restricted Gaussian maximum
 likelihood method for estimating the covariance function parameters
 and the computation of the best unbiased predictor. Our
 approach produces a new set of multi-level contrasts that decouples
 the covariance parameters from the deterministic components. In
 addition, the covariance matrix exhibits fast decay independently
 from the decay rate of the covariance function.  Due to the fast
 decay of the covariance matrix only a small set of coefficients of
 the covariance matrix are computed with a level-dependent criterion.
 We showed results of our method for the Mat\'{e}rn covariance with
 highly irregular placement of the observation locations to good
 accuracy.


We are currently working on deriving error
estimates of the kriging estimate and determinant computation with
respect to the number of degrees of freedom $n$. We are also
contemplating extending our multi-level approach to multivariate
random fields and cokriging (e.g. \cite{Furrer2011}).

Our method also applies to
non-stationary problems if the covariance function is differentiable
to degree $\tilde{f}$. For example, if the covariance function changes
smoothly with respect to the location, Lemma
\ref{multilevelcov:lemma1} still applies and the multi-level
covariance matrix decays at the same rate as a stationary one. Now,
even if the covariance function is non differentiable everywhere with
respect to the location, Lemma \ref{multilevelcov:lemma1} still
applies, but at a lower decay rate.

\corb{Note that we have not made direct comparisons with many of the
  approaches described in Section \ref{Introduction}. These methods
  are very good at solving a particular type of problem i.e. grid-like
  geometries and/or compact covariance functions. For these situations
  we recommend using the approaches already developed in the
  literature as they work well and are easier to implement. However,
  to our knowledge we have no seen fast results where the placement of
  the observations are highly irregular and the Mat\'{e}rn covariance
  function decays slowly. We plan to include some comparison for the
  spatial covariance paper we plan to write.}

\section*{Appendix A: Proofs}

{\bf Lemma \ref{multilevelcov:lemma1}:}

Since $\phi(r;\btheta)$ is in $C^{\tilde{f}+1}(\R)$, then by Taylor's
theorem we have that for every $\bx \in B_{\ba}$
$\phi(\bx,\by;\btheta) = \sum_{|\alpha| \leq {\tilde f}}
\frac{D^{\alpha}_{\bx}\phi(\ba,\by;\btheta)}{\alpha
  !}(\bx-\ba)^{\alpha} + R_{\alpha}(\bx,\by;\btheta)$,
\noindent where $(\bx-\ba)^{\alpha}:=(x_{1}-a_{1})^{\alpha_{1}} \cdots
(x_{d}-a_{d})^{\alpha_{d}}$, $\alpha ! := \alpha_{1}! \cdots
\alpha_{d}!$, and $R_{\alpha}(\bx,\by;\btheta) :=
\sum_{|\alpha|=\tilde{f}+1}\frac{ (\bx-\ba)^{\alpha}}{\alpha!}
D^{\alpha}_{\bx} \phi(\ba+s(\bx-\ba),\by;\btheta)$ for some $s \in
[0,1]$.  Now, recall that $\bpsi^{i,k}_{\tilde{k}}$ is orthogonal to
${\cal P}^{p}(\mathbb{S})$ then $\sum_{h = 1}^{n}
\bpsi^{i,k}_{\tilde{k}}[h] \phi(\bs_h,\by;\btheta) = \sum_{h = 1}^{n}
\bpsi^{i,k}_{\tilde{k}}[h]$ $R_{\alpha}(\bs_h,\by;\btheta)$.  Since
$\bpsi^{j,l}_{\tilde{l}}$ is also orthogonal to ${\cal
  P}^{p}(\mathbb{S})$ then by applying Taylor's theorem centered at
$\bb \in B_{\ba}$:
\[
\begin{split}
&|(\bpsi^{i,k}_{\tilde{k}})\T \bC(\btheta)
\bpsi^{j,l}_{\tilde{l}}|
= 
\left|\sum_{h = 1}^{n} 
\sum_{e = 1}^{n} 
\bpsi^{i,k}_{\tilde{k}}[h] \bpsi^{j,l}_{\tilde{l}}[e]
\phi(\bs_h,\bs_e;\btheta) \right| 
=
\left| \sum_{h = 1}^{n} \sum_{e = 1}^{n} \left(
\sum_{|\alpha|=\tilde{f}+1}\frac{(\bs_h-\ba)^{\alpha}}{\alpha!}  \right. \right. \\
& \left. \left. 
\sum_{|\beta|=\tilde{f}+1}\frac{(\bs_e-\bb)^{\beta}}{\beta!}
D^{\alpha}_{\bx} D^{\beta}_{\by}
\phi(
a+s(\bs_h-\ba), b+t(\bs_e-\bb);\btheta) \bpsi^{i,k}_{\tilde{k}}[h] \bpsi^{j,l}_{\tilde{l}}[e]
\right) \right| \\
\leq&
\sum_{|\alpha|=\tilde{f}+1} \sum_{|\beta|=\tilde{f}+1
}
\frac{r_{\ba}^{\alpha}}{\alpha!} \frac{r_{\bb}^{\beta}}{\beta!}   
|  D^{\alpha}_{\bx} D^{\beta}_{\by} \phi(\ba+s(\bs_h-\ba), \bb+t(\bs_e-\bb);\btheta)|
\left|\sum_{r = 1}^{n} \sum_{e = 1}^{n} 
\bpsi^{i,k}_{\tilde{k}}[r] \bpsi^{j,l}_{\tilde{l}}[e]\right|\\
\leq& 
\sum_{|\alpha|=\tilde{f}+1} \sum_{|\beta|=\tilde{f}+1}
\frac{r_{\ba}^{\alpha}}{\alpha!} \frac{r_{\bb}^{\beta}}{\beta!}   
\sup_{\bx \in B_{\ba}, \by \in B_{\bb}} |D^{\alpha}_{\bx} D^{\beta}_{\by} \phi(\bx, \by;\btheta)|,
\end{split} 
\]
for some $s,t \in [0,1]$. The last inequality follows since from
Schwartz' inequality $\sum_{h = 1}^{n} |\bpsi^{i,k}_{\tilde{k}}[h]|
 \leq
\sqrt{ \sum_{h = 1}^{n} (\bpsi^{i,k}_{\tilde{k}}[h])^2} = 1$ and
$\sum_{e = 1}^{n} |\bpsi^{j,l}_{\tilde{l}}[e]| \leq
\sqrt{ \sum_{e = 1}^{n} (\bpsi^{j,l}_{\tilde{l}}[e])^2} = 1$.
\hfill $\Box$

\noindent 
{\bf Lemma \ref{multilevelkriging:lemma2}:} Since the Mat\'{e}rn
covariance function $\phi(\bx,\by; \btheta)$ is positive definite
we have that for all $\bv \neq {\bf 0}$:
\[
\sum_{i,j=1}^{n} v_i v_{j} \bC^{i,j}(\btheta) = 
\sum_{i,j=1}^{n} v_i v_{j} \phi(\bx_{i},\by_{j}; \btheta) > 0,
\]
where $\bC^{i,j}(\btheta)$ is the $(i,j)$ entry of the matrix
$\bC(\btheta)$.  The diagonal terms of $\bC_{W}$ are of the form
$(\bpsi^{i,k}_{\tilde{k}})\T \bC(\btheta) \bpsi^{i,k}_{\tilde{k}} $.
This implies that

\[
(\bpsi^{i,k}_{\tilde{k}})\T \bC(\btheta)
\bpsi^{i,k}_{\tilde{k}}  > 0.
\]
Thus, $\bD_W$ will always be positive definite.
\hfill $\Box$

\section*{Appendix B: Notation}

{\footnotesize
\corb{
\begin{tabular}{l l  l l}
\multicolumn{4}{l}{\bf Index for when the following are first defined, 
mentioned or reformulated.} \\
$d \in \mathbb{N}$               & Dimension of problem (p1)&
$n \in \mathbb{N}$               & Number of observations (p1) \\
$t \in \mathbb{N}$               & Maximum MB level (p8) &
$p \in \mathbb{N}$               & Number of columns of $\bM$ (p1) \\
$f \in \mathbb{N}$               & Polynomial degree (p7) &
$\tilde{p} \in \mathbb{N}$       & Accuracy parameter of MB (p7) \\
$\tilde{f} \in \mathbb{N}$       & Degree of multilevel basis. p(7) &
$w \in \mathbb{N}$               & Dimension of $\btheta$  (p1) \\
$\bM \in \R^{n \times p}$           & Design matrix (p1) &
$\phi(\cdot)$                    & Covariance function (p7) \\
$\mathbb{S} := \{ \bs_1,\dots,\bs_n\}$    & Locations of observations (p1) &
$\btheta:=(\nu,\rho) \in \R^{2}$  & Param. of matern kernel (p1) \\
$\bC(\btheta) \in \R^{n \times n}$  & Covariance matrix (p1) &
$\bZ \in \R^{n}$                  & Observation values (p1) \\
$\bs_{0} \in \R^{d}$               & Target points set (p1) &   
${\cal P}^{p}(\mathbb{S})$        & Span of the columns of  $\bM$ (p1) \\    
$\bbeta \in \R^{p}$               & Vector of unknowns from & 
$\hat{\bbeta} \in \R^p $,        & Estimates \\
                                 & deterministic model (p1) &   
$\hat{\bZ}(\bs_0)$               & kriging estimate at $\bs_0$ (p2) \\
$l(\bbeta,\btheta) \in \R$       & Log-likelihood function (p1) &
$B^{q}_k$                         & Cube at level $q$ and index $k$ (p8)\\
$\bL$                           &  $\R^n \rightarrow {\cal P}^{p}(\mathbb{S})$ (p4) & 
$\bW$                           &  $\R^n \rightarrow ({\cal P}^{p}(\mathbb{S}))^{\perp}$ (p4) \\
$\bP \T$                           &  
$ \left[
\begin{array}{cc}
\bW \T &
\bL \T
\end{array}
\right
]$ (p10) & $\tilde{\bZ}^i_W$ &  (p17) \\
                      & 
    &
$\tilde{\ell}^i_{W}(\btheta)$        & Multilevel log-likelihood (p17)\\
${\cal Q}^d_f$ & Set of polynomial monomials & 
$\tau \in \mathbb{N}_0$  & Level dependent \\
               & of order $f$ and dimension $d$ (p7)& & criterion constant (p13)\\
\end{tabular}
}
}
\section*{Appendix C}

Using the approach developed in this paper we can
compute the MSE at the target point $\bs_0$. Now, since $\bP\T \bP =
\bI$ we have that $\bM_f \T \bC(\btheta)^{-1} \bM_f = \bM_f \T \bP\T
\bP \bC(\btheta)^{-1} \bP \T \bP \bM_f = \bM_f \T \bP\T
\bK_{W}(\btheta)^{-1}\bP \bM_f$, $\bK_W(\btheta) := \bP \bK(\btheta)
\bP \T$, $\bK_{W}(\btheta):= \left[
\begin{array}{c c}
 \bC_W(\btheta) & \ba_W \\
\ba_W \T & \bS_W(\btheta)
\end{array}
\right ]$ 
where $\bS_W(\btheta) := \bL \bC(\btheta) \bL \T \in \R^{p
  \times p}$ and $\ba_W(\btheta) := \bW \bC(\btheta) \bL \T \in \R^{
  (n - p) \times p}$. Let
$\tilde{\bS}_W(\btheta):=(\bS_W(\btheta) - \ba_W \T \bC_W(\btheta)^{-1} \ba_W)^{-1}$,
then
\[
\bK_{W}(\btheta)^{-1}:= \left[
\begin{array}{c c}
\bC_W(\btheta)^{-1}  +
\bC_W(\btheta)^{-1} \ba_W \tilde{\bS}_W(\btheta) \ba_W \T \bC_W(\btheta)^{-1}
& -\bC_W(\btheta)^{-1} \ba_W \tilde{\bS}_W(\btheta) \\
-\tilde{\bS}_W(\btheta) \ba_W\T \bC_W(\btheta)^{-1}
& \tilde{\bS}_W(\btheta) 
\end{array}
\right].
\]
Given that $\bW \bM_f  = {\bf 0}$ then $ \bM_f \T \bC(\btheta)^{-1} \bM_f
= (\bL \bM_f)^T \tilde{\bS}_W(\btheta) (\bL \bM_f) $.  Following a
similar argument we have that equation \eqref{Kriging:MSE} becomes
\begin{equation}
1 + 
\tilde{\bu}
\T
\bM_f \T 
(\bL \bM_f) \T \tilde{\bS}_W(\btheta) (\bL \bM_f)
\tilde{\bu}
-\bc_{W}\T\bK_{W}^{-1}(\btheta)\bc_{W},
\label{Kriging:MSE2}
\end{equation}
where $\bc_W := \bP \bc$.  By using matrix-vector products with the
PCG method
 $\tilde{\bS}_W(\btheta)$ can be computed in ${\cal O}((k(t+2)+1)np +
p^d)$. Thus the term $(\bL \bM_f) \T \tilde{\bS}_W(\btheta) (\bL
\bM_f) $ can be computed in ${\cal O}((k(t+2)+1)np + p^d + (t+2)pn)$.  Now, by
also using matrix-vector products the term
$\bK_{W}^{-1}(\btheta)\bc_{W}$ can be computed in ${\cal O}(kn(t+2) + p^d
)$. Thus the total cost for computing equation \eqref{Kriging:MSE2} is
${\cal O}((k(t+2)+1)np + p^d + (t+2)pn)$.\\

\noindent {\large {\bf Acknowledgements:}} We appreciate the help and
advice from Jun Li and Lisandro Dalcin in getting the C++ code working
properly, and to Stefano Castruccio for giving us feedback on our
manuscript.

\baselineskip 22pt
\pagestyle{empty}

\clearpage
\pagenumbering{gobble} 
\end{document}